\documentclass[prb,twocolumn,aps,superscriptaddress]{revtex4-1}
\usepackage{graphicx}
\usepackage{amsmath}
\usepackage[colorlinks=true, citecolor=blue, urlcolor=blue, linkcolor = blue]{hyperref}
\usepackage{amssymb}
\usepackage{bm}
\usepackage{hyperref}
\usepackage{cancel}
\usepackage[normalem]{ulem} 
\usepackage{soul} 
\usepackage[dvipsnames]{xcolor}
\usepackage{float} 
\usepackage[caption=false]{subfig}
\usepackage{mathtools} 
\definecolor{lgray}{gray}{0.8}

\newcommand{\TZtwoDW}{1130 U_0} %
\newcommand{\TZtwoColl}{790 U_0}

\begin{document}

\title{Crystalline droplets with emergent color charge \\in many-body systems with sign-changing interactions}
\author{P. Karpov}
\email{karpov@pks.mpg.de}
\affiliation{Max Planck Institute for the Physics of Complex Systems, N{\"o}thnitzer Stra{\ss}e 38, Dresden 01187, Germany}
\affiliation{National University of Science and Technology ``MISiS'', Moscow, Russia}
\author{F. Piazza}
\affiliation{Max Planck Institute for the Physics of Complex Systems, N{\"o}thnitzer Stra{\ss}e 38, Dresden 01187, Germany}

\date{November 24, 2019}

\begin{abstract}

 We introduce a novel type of self-bound droplet which carries an emergent color charge. We consider a system of particles hopping on a lattice and interacting via a commensurately sign-changing potential which is attractive at a short range.
The droplet formation is heralded by spontaneous crystallization into topologically distinct domains. 
This endows each droplet with an emergent color charge governing their mutual interactions: attractive for equal colors and repulsive otherwise. The number of allowed colors is fixed only by the discrete spatial symmetries of the sign-changing part of the interaction potential.
With increasing interaction range, the droplets become progressively more mobile, with their color charge still being energetically protected, allowing for nontrivial viscous dynamics of the interacting droplet plasmas formed during cooling.
Sign-changing potentials with a short-range attraction appear quite naturally for light-mediated interactions and we concretely propose a realization in state-of-the-art experiments with cold atoms in a multimode optical cavity.

\end{abstract}

\maketitle


{\it Introduction.}
Classical and quantum many-body systems often feature competitions between attractive and repulsive forces, including kinetic terms such as thermodynamic or quantum pressure. Attractive forces favor the formation of clusters of matter, while repulsive and kinetic terms prevent the collapse of the system. Through competition between attraction and repulsion various self-bound objects may emerge: from atomic nuclei \cite{Bender:2003}, van der Waals nanoclusters \cite{Chalasinski:1994}, droplets in conventional liquids \cite{Frohn:2000-book} and in liquid helium \cite{Barranco:2006}, to stars, galaxies \cite{Padmanabhan:1993-book}, and black holes \cite{Dvali:2014}.
(Ultra)cold-atomic systems represent a convenient playground for studying such self-bound objects. {Quantum droplets were recently observed in bosonic dipolar gases \cite{Pfau:2016-Nature,Chomaz:2016} and in bosonic mixtures \cite{Tarruell:2018-Science,Semeghini:2018}.} 

Here, we introduce another type of droplet with the distinguishing property of carrying an emergent color charge of topological origin. 
We show that such droplets can form and be stably self-bound if the interparticle interactions are (i) sign changing in space commensurately with the lattice, (ii) attractive on the same lattice site, and (iii) finite-ranged. Features (ii) and (iii) allow for the formation of droplets which are self-bound below a certain critical temperature. Crucially, it is feature (i) which endows the droplets with the charge, as they can then only be stable if their constituent particles occupy a specific sublattice, i.e., take a specific ``color''. The latter indeed determines the sign of the interaction between droplets: attractive for the same color and repulsive otherwise. The integer number of allowed colors thus depends solely on the discrete spatial symmetries of the lattice and the interaction potential. We find that the color charge of a droplet is conserved due to an energetic protection, as it would take an extensive number of local single-particle processes to remove it. Moreover, each of those processes is hindered by an energy barrier which also scales extensively in the thermodynamic limit. 

The emergence of a droplet color charge is intimately connected with the phenomenon of the spontaneous breaking of discrete spatial translational invariance by choosing a specific sublattice. Indeed, we observe that droplet formation is necessarily heralded by crystallization: upon cooling down the system, crystalline domains are formed, and each one then subsequently condenses into one or several droplets with the same color charge. The following cooling dynamics is one of a viscous plasma of droplets with different color charges governing their mutual interactions, whose range is set by the microscopic interaction range.
For sufficiently long interaction ranges, we also observe stable, translationally invariant crystalline phases at intermediate temperatures. 

While the microscopic features (i) and (ii) might seem exotic in the context of condensed-matter physics, they are quite generically found in systems of particles with light-mediated interactions, provided that specific electromagnetic modes are selected via laser driving and/or light confinement. This is, for instance, the case for experiments involving laser driven-atoms in optical resonators \cite{Ritsch:2013,Black:2003,Baumann:2010,Arnold:2012-PRL,Hemmerich:2014,Leonard:2017,Vaidya:2018,Lev:2019-short,Lev:2019-long} or photonic crystals \cite{Tudela:2015}. In particular, the formation of noncharged droplets due to light-mediated interactions in a noncommensurate case has been predicted recently \cite{Pohl:2018}.


\begin{figure}[t] %
\centering
\includegraphics{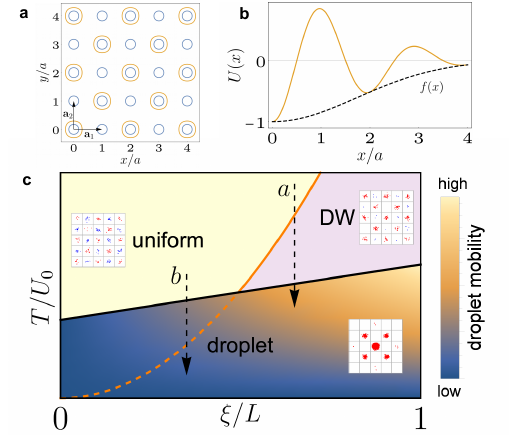}
\caption{ 
        (a) Contours encircling the local minima of the interaction potential (orange color) and of the external lattice (blue color). 
        (b) Two- and three-dimensional plots of the interaction potential $U$.
        (c) Qualitative equilibrium phase diagram in coordinate temperature $T/U_0$ vs interaction range $\xi/L$, obtained from eqs. (\ref{eq:T_CDW-arbitrary-D-N}) and (\ref{eq:Tcoll-arb_D}). The black line denotes the droplet-formation phase transition while the orange line the DW phase transition. The letters {\it a} and {\it b} indicate cuts through the phase diagram with and without an intermediate DW phase, corresponding to Fig. \ref{fig:Z2_modeling_plots}a, b.  
}
\label{fig:potential_parallel_perpendicular_polarizations}
\label{fig:phase_diagram}
\end{figure}

{\it Model.}
We consider particles under the influence of the following potential,
$H = \sum_{k} V(\mathbf{x}_k) +\sum_{k<l} U(\mathbf{x}_k, \mathbf{x}_l)$.
For a deep external potential $|V| \gg |U|$ we can consider a simplified lattice version of the model.
Provided that the interaction potential $U$ satisfies the properties (i)-(iii) listed in the Introduction, the phenomenology we are going to describe further depends solely on the spatial symmetry of the interaction potential relative to the external lattice. Being in a commensurate case, the above symmetry would be the one being spontaneously broken by entering a crystalline, specifically a density-wave (DW), phase. 
This discrete symmetry will also determine the set of possible colors of the droplet charge, as we shall see below. 
For concreteness, we will restrict ourselves to a $\mathbb{Z}_2$ symmetry emerging in the two-dimensional configuration illustrated in Fig.~\ref{fig:potential_parallel_perpendicular_polarizations}.
We parametrize the interaction potential between two particles, one of which is situated at the origin and another at site $\mathbf{r}_{i,j} = i\mathbf{a}_1+ j \mathbf{a}_2$, as  $U_{i, j} \equiv U(0,\mathbf{r}_{i,j}) = (-1)^{i+j} f(r_{i,j})$ (where $\mathbf{a}_1, \mathbf{a}_2$ are the basis vectors of the lattice).
The envelope function $f(r)<0$ is monotonously increasing up to zero with some characteristic range $\xi$.
In the following, we consider a thermal regime where the dynamics is classical (the regime of applicability is discussed in \cite{SM}). We performed Metropolis Monte Carlo (MC) simulations of the discrete two-dimensional model with a particular choice of the interaction potential $U_{i,j} = -U_0 (-1)^{i+j} \exp(-(i^2+j^2)/\xi^2)$, which possesses the generic properties described above  (we use the lattice constant $a$ as the unit of length).
It is crucial to note that it is the sign-changing part of the interaction potential $(-1)^{i+j}$ that possesses the aforementioned $\mathbb{Z}_2$ symmetry defining the two possible types of color charges.

{\it Thermodynamics.}
The qualitative form of the equilibrium phase diagram of our model, shown in Fig.~\ref{fig:phase_diagram}, can be simply understood by considering the free energy, which also provides analytical estimates for the transition temperatures.
At high temperatures we expect all the particles to be uniformly spread over all sites of the lattice. By decreasing the temperature, the sign-alternating nature of the interaction potential favors a checkerboard DW state, while the $\xi$-range attraction favors a droplet state with all the particles concentrated within one cell of size set by the range $\xi \times \xi$ of the interaction potential (note that while at $T\rightarrow0$ we expect a droplet to be localized to just one site, but for nonzero temperatures, the droplet can be smeared up to size $\xi$ by the entropic force).
Let us compare the free energies of these three states: uniform, DW, and droplet. For generality we consider a $D$-dimensional case, when the particles sit in a box with volume $L^D$. In the uniform state, all the sites are populated by an approximately equal number of particles $n \equiv N/L^D$. Each particle gives a contribution $\frac12 (-U_A+U_B)n$ to the energy, where $-U_{A} = -\sum_{i+j=\mathrm{even}} |U_{i, j}|$ is the negative (attractive) part coming from the interaction with all particles sharing the same sublattice, while $U_B = \sum_{i+j=\mathrm{odd}} U_{i,j}$ is the positive (repulsive) part from the other sublattice. The entropy of the state is $S_{\mathrm{unif}}  \approx N\ln L^D$, resulting in a free energy of the uniform state given by
$
F_{\mathrm{unif}} = -\frac{1}{2} (U_A-U_B) n^2 L^D - T N \ln L^D.
$
In the checkerboard DW state, all the particles occupy one sublattice with $2n$ particles per site on average. The free energy of the DW state is thus
$
F_{\mathrm{DW}} = -\frac{1}{2} U_A (2n)^2 \frac{L^D}{2} - TN \ln (L^D/2)\ .
$
In the droplet state, all particles occupy a given sublattice within a box of size $\xi^D$ and each of them interacts with all the others with a corresponding energy of the order of $-U_0 \equiv U(0)$, so that the free energy can be estimated as
$
F_{\mathrm{dr}} \approx -\frac{1}{2} U_0 N^2-T N \ln \xi^D.
$
In the particular case of a short-range potential, $\xi \lesssim 1$, all the atoms would collapse to one lattice site.

The droplet state is the lowest energy one and thus the most favorable at low $T$. 
Hence there are two scenarios for phase transitions with lowering the temperature: (a) transition from the high-temperature uniform state to the DW state (and then to the droplet state at some lower temperature); (b) transition from the uniform state directly to the droplet state.
For case (a), from $F_{\mathrm{unif}} = F_{\mathrm{DW}}$ we get the critical temperature
\begin{align}
T_{\mathrm{DW}} =\frac{N (U_A +U_B)}{L^D\cdot 2 \ln 2}\sim \frac{N U_0 \xi^D}{L^D}\,
\label{eq:T_CDW-arbitrary-D-N}
\end{align}
where we have used the identity $C_D \xi^D U_0 \equiv U_A+U_B = \sum_{i, j} |U_{i,j}|$, with $C_D$ being the dimensionless factor for the volume of a $D$-dimensional sphere with radius $r$: $V = C_D r^D$. 
Using mean-field theory it is possible to quantitatively determine $T_{\mathrm{DW}}$ [\onlinecite{SM}]. 
For case (b), from $F_{\mathrm{unif}} = F_{\mathrm{dr}}$  we get
\begin{align}
T_{\mathrm{dr}}  =  \frac{N U_0}{2\ln (L/\xi)^D}
\label{eq:Tcoll-arb_D}
\end{align}
Comparing expressions (\ref{eq:T_CDW-arbitrary-D-N}) and (\ref{eq:Tcoll-arb_D}) we see that if $\xi=\text{const}$ in the thermodynamic limit $N,L \rightarrow \infty$ (while $n=N/L^D=\text{const.}$) we always have $T_{\mathrm{dr}}>T_{\mathrm{DW}}$, so that the droplet phase is always more favorable than the DW phase.
However, if the range of the potential also scales with the size of the system as $\xi \sim L$, then both transition temperatures scale identically as $L\rightarrow\infty$, which allows for the existence of the DW phase at intermediate temperatures even in the thermodynamic limit. In \cite{SM}  we present an exactly solvable toy model for the droplet-formation phase transition. In general, the qualitative dependence of the critical temperatures on the interaction range is shown by the black and orange lines in the phase diagram of Fig. \ref{fig:phase_diagram}. 

\begin{figure}[t] 
\centering
\includegraphics[]{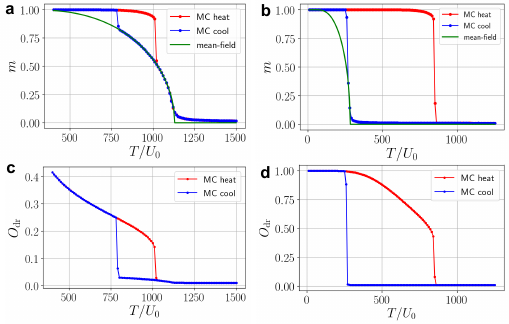}
\caption{Temperature dependence of the order parameters for $\xi=6$ (left column) and $\xi=3$ (right column). (a),(b) Sublattice imbalance order parameter. (c),(d) Droplet order parameter. The results shown here and below were obtained using a lattice of size $L\times L=30\times 30$ filled with $N=9\times10^3$ particles, 
a number of MC steps $N_{\mathrm{steps}}=4.5 \times10^7$, and of MC sweeps $N_{\mathrm{sweeps}}=N_{\mathrm{steps}}/N = 5\times10^4$ at a given temperature.
}
\label{fig:Z2_modeling_plots}
\end{figure}

\begin{figure}[!t] 
\centering
\includegraphics{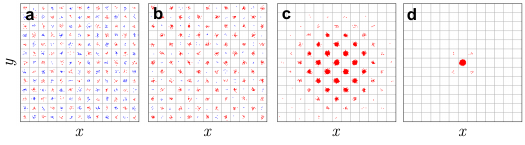}
\caption{Snapshots of the size $13\times13$ of the system $30\times30$, $\xi=6$, at different temperatures. 
	(a) Uniform state: $T=1200U_0 > T_{\mathrm{DW}}$. 
	(b) Density-wave state: $T_{\mathrm{dr}}<T=800U_0<T_{\mathrm{DW}}$. 
	(c) Droplet state: $T=700U_0 < T_{\mathrm{dr}}$. 
	(d) Droplet collapsed to a single site: $T=50U_0 \ll T_{\mathrm{dr}}$, this state is connected to the extended droplet state by a crossover. Particles are colored in red and blue according to the checkerboard sublattice they occupy. All positions of particles inside a given cell are equivalent -- small random noise in the position was added only for a better visual representation.}
\label{fig:Z2_modeling_positions}
\end{figure}

Our MC simulations for the case $T_{\mathrm{dr}}<T_{\mathrm{DW}}$ confirm the existence of a stable DW phase. In order to trace the DW transition we use the even-odd sublattice imbalance order parameter $m=(N_{\mathrm{even}}-N_{\mathrm{odd}})/N$; for the droplet-formation phase transition we use the order parameter related to breaking of translational symmetry of the system
$
O_{\mathrm{dr}} = \sqrt{\frac{L^2}{L^2-1} \sum_i \left(\frac{n_i}{N}-\frac{1}{L^2} \right)^2},
$
so that $O_{\mathrm{dr}}=0$ corresponds to the uniform phase and $O_{\mathrm{dr}}=1$ to the single-site droplet.
For $\xi=6$ (corresponding to the dashed arrow labeled by $a$ in Fig.~\ref{fig:phase_diagram}), we indeed observe both phase transitions with lowering the temperature:
a continuous transition from the uniform to the DW phase at $T\approx \TZtwoDW$ and 
 a first-order phase transition to the droplet phase at lower $T\approx \TZtwoColl$ on cooling and $T\approx1010 U_0$ on heating  (Fig. \ref{fig:Z2_modeling_plots}).
Figure \ref{fig:Z2_modeling_positions} directly visualizes the sequence of the phase transitions
showing the snapshots of the system at different temperatures. The initial droplet of size $\xi \times \xi$ (Fig. \ref{fig:Z2_modeling_positions}c) with lowering the temperature undergoes a crossover to the state where it becomes localized on one site (Fig. \ref{fig:Z2_modeling_positions}d).
Note that even such a localized droplet has a ``memory'' about the sublattice of the intermediate DW state it originates from. 

The situation is more complex in the case $T_{\mathrm{dr}}>T_{\mathrm{DW}}$, where the free-energy arguments above however do not provide the full picture. There we have assumed one crucial feature of the droplet, namely that it occupies a single sublattice, as confirmed already by the numerical simulations shown in Fig.~\ref{fig:Z2_modeling_positions}. This is due to the fact that only such a droplet is energetically favorable against the uniform phase because of the sign-changing nature of the interactions. Droplets in our system have to choose a sublattice, which demonstrates a deep interconnection between the formation of a DW and droplets. This has important implications both for thermodynamics and dynamics. Deferring the discussion of the cooling dynamics to the next section, we report here the result of our MC simulations for the critical temperature of droplet formation out of the homogeneous phase for the case $T_{\mathrm{dr}}>T_{\mathrm{DW}}$. We observe that on cooling the critical temperature is not given by our estimated $T_{\mathrm{dr}}$, but rather by the DW-critical temperature $T_{\mathrm{DW}}$. In the temperature region $T_{\mathrm{DW}}<T<T_{\mathrm{dr}}$ the uniform phase is still metastable and we observe hysteresis (see Fig. \ref{fig:Z2_modeling_plots}c,d and dashed orange line in Fig.~\ref{fig:phase_diagram}). This is because during cooling the long-wavelength density fluctuations do not gain energy (while losing the entropy) due to the sign-changing nature of the interaction potential.
On the contrary, below $T_{\mathrm{DW}}$ one sublattice is spontaneously chosen so that long-wavelength density fluctuations gain energy and allow for the formation of the droplets. During cooling the second-order DW-transition thus triggers the first-order droplet transition. This picture will be confirmed by our simulations of the cooling dynamics that we discuss next.


\begin{figure}[!tbh] 
	\centering
	\includegraphics{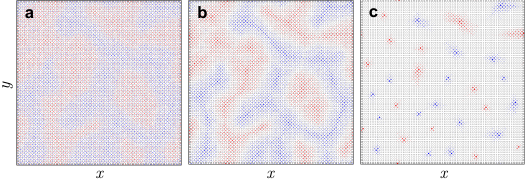}
	\caption{Monte Carlo evolution of the $100\times100$ system with $\xi=5$, instantaneously quenched from the infinite temperature to $T=U_0$. Red and blue dots represent atoms at even and odd sublattices, respectively. 
		(a) DW state with several domains (after three MC sweeps). 
		(b) Domains become sharply separated by regions without particles (after ten MC sweeps). 
		(c) Droplet phase with even- and odd-sublattice types of droplets (after 50 MC sweeps). 
	}
	\label{fig:quench}
\end{figure}

{\it Cooling dynamics.}
The fact that droplet formation is heralded by crystallization into a DW where particles spontaneously choose a sublattice effectively endows droplets with an emergent charge (set by the choice of sublattice) governing their mutual interaction. The latter has a range $\xi$ and, as a consequence of the commensurate sign-changing nature, is attractive between the similar-color charges and repulsive otherwise (see \cite{SM}). The discrete number of possible charge colors is determined by the number of distinct sublattices, i.e., by the discrete spatial symmetry which is spontaneously broken by the DW. Despite our specific example involving a $\mathbb{Z}_2$ symmetry, i.e., two charge types, other numbers of such ``color'' charges can be implemented using appropriate geometries (see  \cite{SM} for the discussion of the four-color case).
The set of droplet colors has a topological nature in the same sense domain-wall defects in a DW state do. 
For the latter, however, only the domain wall separating two infinite domains is topologically protected as it cannot be removed by local perturbations. For finite domains the protection is reduced to a weaker energetic one -- there is an extensive energetic barrier for shifting a domain to a different sublattice. In the same fashion, our color-charged droplets are local objects that can not be topologically protected but are rather energetically protected by an extensive energetic barrier, as we discuss below.

 We consider the case $T_{\mathrm{DW}}<T_{\mathrm{dr}}$ (for $\xi=5$, $L=100$) and use a rapid quench from a random configuration to a low temperature $T\ll T_{\mathrm{DW}}$, tracing then the MC evolution of the state at this fixed temperature (Fig. \ref{fig:quench}).
 Within several MC sweeps particles form domains corresponding to even and odd sublattices (Fig. \ref{fig:quench}a). Because of the repulsion between particles occupying different sublattices, the domains separate from each other with the formation of intermediate low-density regions of width $\sim\xi$ (Fig. \ref{fig:quench}b). After the formation of domains, the attractive interaction between particles within a domain becomes efficient and induces self-collapse. Each domain, depending on its size, decomposes into one or several droplets, preserving the $\mathbb{Z}_2$-color charge of the initial domain (Fig. \ref{fig:quench}c). 

The droplets can however perform a diffusive motion, with the diffusion constant being determined by $T$ and the interaction range $\xi$.
For droplets collapsed to a single site (i.e., at $T\ll T_{\mathrm{dr}}$), the particle-by-particle flow of the droplet becomes suppressed by an energy barrier $\sim  (U_{1,1}-U_0) N_0^2$ (where $N_0$ is the number of atoms in the droplet). However, for droplets localized in a $\xi \times \xi$ region (i.e., at $T\lesssim T_{\mathrm{dr}}$) the barrier is estimated to be not higher than $\sim (U_{1,1}-U_0) (N_0/\xi^2)^2 \sim U_0 N_0^2 /\xi^6$, so that with increasing $\xi$ the movement of the droplets becomes possible, while the barrier  $\sim  U_0 N_0^2$ still protects the total color charge of the droplets of each species. This protection is a required feature for an emergent charge: it needs indeed to be conserved during the dynamics.  
As anticipated, the protection in our case is of an energetic nature since, as long as $N_0$ is extensive, there are required $N_0$ local single-particle steps, each of which is suppressed by an energy barrier scaling itself with $N_0$.  
By tuning $\xi$ we can find a temperature interval where the droplets are mobile (i.e., can perform a diffusive motion), but with the total color charge still being conserved. Figure \ref{fig:phase_diagram} qualitatively shows the mobility of the droplets in the region of their stability in the phase diagram.

During the cooling and after droplet formation we thus deal with an interesting emergent physics: an overall neutral (apart from statistical fluctuations) plasma of color-charged droplets performing viscous movement and interacting over a range $\xi$. Despite the fact that the true thermodynamically stable state is the single droplet discussed in the previous section, the droplet plasma eventually reaches a metastable state with an extensive lifetime. As droplets with equal color charges merge with one another the plasma will be eventually composed of two droplets (one for each color). Since the remaining droplets repel each other, an extensive energy barrier prevents the system from reaching the single-droplet state.

{\it Implementation with light-mediated interactions.}
In the regime of strong coupling where the back-action between light and matter cannot be neglected, photons effectively mediate interactions between particles.
Sign-changing interaction potentials appear naturally if specific electromagnetic modes are selected via driving and/or confinement.
For example, consider an ensemble of neutral atoms inside an optical cavity \cite{Ritsch:2013}. Let us assume the atomic motion being confined to the $x-y$ plane transverse to the cavity axis and driving a given atomic transition by two off-resonant laser beams forming standing waves, so that the laser field can be written as
$\Omega(\mathbf{x}) = \Omega_0 (\cos(2\pi \tilde{x}/\lambda)+ \cos(2\pi \tilde{y}/\lambda))$, where $\tilde{x}=(x+y)/\sqrt{2}$, $\tilde{y}=(x-y)/\sqrt{2}$, and $\lambda$ is  the pump wavelength which settles the lattice constant to be $a=\lambda/\sqrt{2}$. For red detuning from the atomic transition this creates a two-dimensional optical square lattice potential $V(\mathbf{x}) = -V_0 \Omega^2 (\mathbf{x})$ (Fig. \ref{fig:potential_parallel_perpendicular_polarizations}; see also \cite{SM} for the details of the microscopic model).
In addition, two-photon transitions involving one laser- and one cavity-photon generate an interaction potential between the atoms which takes the form $U(\mathbf{x},\mathbf{x}') = \Omega(\mathbf{x}) \Omega(\mathbf{x}')  f(|\mathbf{x}-\mathbf{x'}|)$ [\onlinecite{Ritsch:2013}], where the envelope function $f(r)$ is determined by the geometry of the cavity. For a near-planar cavity with only one mode close to resonance, we have $f(r) = const$, i.e., a global interaction with $\xi=\infty$, while for an infinite number of resonant modes (for example, a concentric or confocal cavity) we have $f(r) =- \delta(r)$ (completely localized interaction). For a realistic case of large but finite number of nearly degenerate modes realized recently in Stanford \cite{Vaidya:2018}, $f(r)<0$ is typically a monotonously increasing up to zero function with some characteristic range $\xi$, exactly as we have considered in the present work.
A modification of the described scheme using pump lasers with orthogonal polarizations also allows us to realize a more complex case of four color-charge species (see  \cite{SM} for the details).

DW phases of atoms in optical cavities so far have been experimentally studied in the regime where the interaction range $\xi$ is larger than the cloud's size \cite{Black:2003,Baumann:2010,Arnold:2012-PRL,Hemmerich:2014,Esslinger:2016,Kollar:2017,Leonard:2017}. Only recent experimental developments using multimode cavities \cite{Vaidya:2018,Lev:2019-short,Lev:2019-long}  have demonstrated the tunability of the range down to scales below typical cloud's sizes, at least for the thermal case.
Here, we have shown that sign-changing interactions behave drastically different as soon as the finiteness of the interaction range is appreciable, offering generic scenarios for the phenomenon of crystallization which have not been considered so far.
The proposed scheme should be feasible within the current experimental capabilities: in particular, 
the magnitude and the color charge of the droplets can be nondestructively monitored by imaging the amplitude and the phase of the scattered light \cite{Lev:2019-short,Lev:2019-long}. 
Since the droplets exist in the classical (thermal) regime, it is possible to work in a much larger temperature range and with much bigger atomic clouds, so that the interaction range is smaller than the cloud’s size.
For a typical interaction scale $U_0$ of the order of kHz, the relevant gas temperatures can be estimated as $T\simeq 10^3$kHz $\approx 50\mu$K. No fine tuning is needed to observe the droplet plasma phase.

Finally, we emphasize the limits of applicability of the studied model. The approximation of the lattice gas is applicable when $T\ll V_0$, in this case the atoms' positions in optical lattice cells are bound by a sphere with characteristic radius $r_T \simeq a \sqrt{T/V_0} \ll a$. The quantum effects can be neglected if $n \lambda_T^3 \ll r_T^3$, where $n$ is the number of atoms in one cell and $\lambda_T= (2\pi\hbar^2/m T)^{3/2}$ is the thermal de Broglie wavelength (for simplicity we take the frequency of the potential, confining atoms to the $z=0$ plane to be the same as for the in-plane optical lattice). If $n>n_{BEC} \sim T\sqrt{ma^2/\hbar^2V_0}$ then the nucleation of a droplet is accompanied by its Bose-Einstein condensation.
Contact repulsion of atoms qualitatively preserves the described picture. When the droplet formation begins, every optical lattice site occupied of the forming droplet is being filled up to some maximum density of particles until the contact hard-core repulsion makes unfavorable the further increase of the number of atoms at the site. If $a_{\mathrm{hard}}$ is the hard-core radius of the atoms this happens when the number of atoms  in optical lattice site exceeds $n_{\mathrm{hard}} \sim (r_T/a_{\mathrm{hard}})^3 = (a/a_{\mathrm{hard}})^3 (T/V_0)^{3/2}.$
Such a contact interaction thus will stabilize the extended droplets with respect to the single-site ones.


{\it Summary and outlook.} 
We introduced a novel type of droplet with the distinguishing property of an emergent color charge. The latter appears in lattice systems with sign-changing interactions (e.g., in current experiments with atoms in multimode cavities), where the phenomenon of crystallization becomes deeply connected with the droplet formation. Future explorations shall involve the study of the quantum regime and in particular the interplay of zero-point energy, particle statistics, as well as possible short-range repulsive potentials with the sign-changing interactions considered here.

 {\it Acknowledgments.} We are grateful to Julian Leonard, Roderich Moessner, Farokh Mivehvar, Helmut Ritsch, and Georgi Dvali for helpful discussions. PK acknowledges the support of the Alexander von Humboldt Foundation and the Ministry of Science and Higher Education of the Russian Federation.


\end{document}


\title{Supplemental Material for ``Crystalline droplets with emergent topological color-charge in many-body systems with sign-changing interactions''}
\author{P. Karpov}
\email{karpov@pks.mpg.de}
\affiliation{Max Planck Institute for the Physics of Complex Systems, N{\"o}thnitzer Stra{\ss}e 38, Dresden 01187, Germany}
\affiliation{National University of Science and Technology ``MISiS'', Moscow, Russia}
\author{F. Piazza}
\affiliation{Max Planck Institute for the Physics of Complex Systems, N{\"o}thnitzer Stra{\ss}e 38, Dresden 01187, Germany}

\date{September 30, 2019}

\maketitle

\onecolumngrid


\section{Density-wave phase transition: mean-field theory}
\label{Appendix Density-wave transition}

In this Supplementary Section we construct a mean-field theory for density-wave phase transition and find the temperature dependence of the order parameter. This can be done if the number of atoms inside a $D$-dimensional sphere with radius $\xi$ is much greater than one: $n \xi^D \gg 1$.

Let $n_A$ and $n_B$ be the average concentrations of particles per cite at the two checkerboard sublattices $A$ and $B$. We define the sublattice imbalance order parameter as
\begin{align}
m(T)= \frac{n_A-n_B}{2n}
\label{eq:order_param_Z2}
\end{align}
Recall that $n = N/L^D$ is average number of particles per site in the uniform phase. We start from the density-wave phase $n_A\approx 2n$, $n_B \approx 0$, so $m\approx 1$ and want to describe a transition to the uniform state $n_A\approx n_B\approx n$, $m\approx 0$.

\begin{figure}[tbh] %
	\centering
	\centering
	\includegraphics[width=0.45\linewidth]{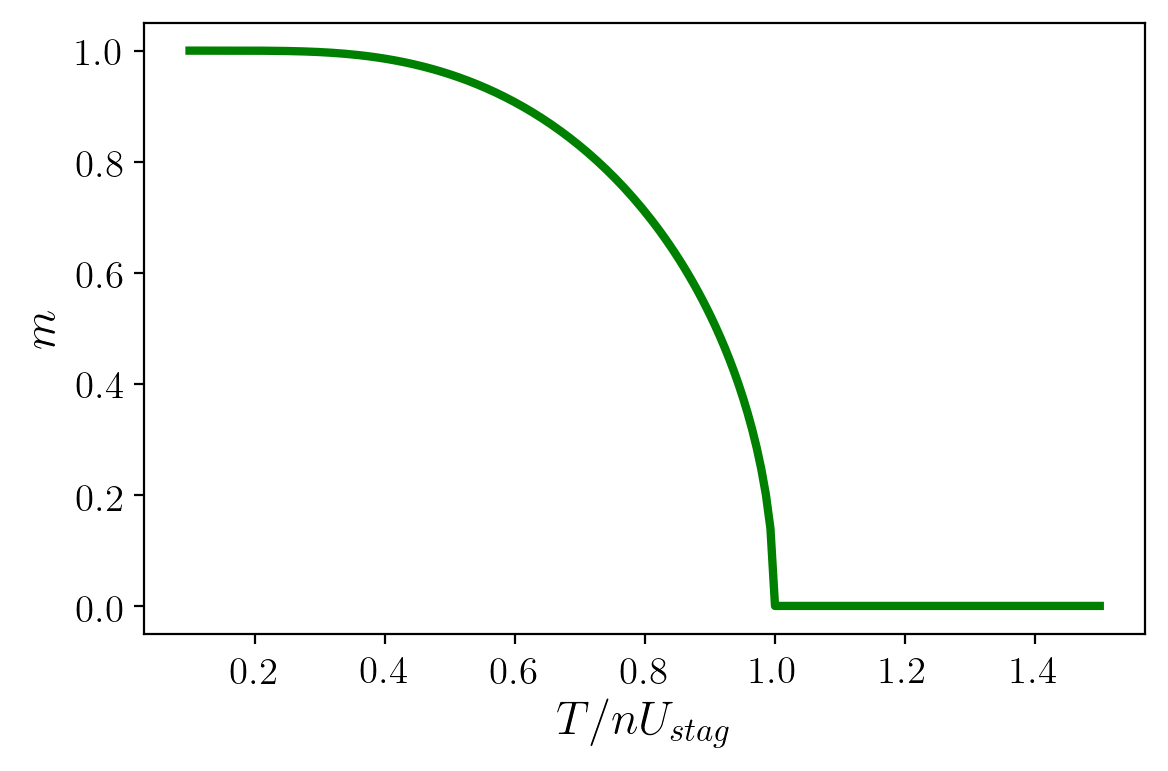}	
	\caption{Analytical mean-field temperature dependence of the sublattice imbalance order parameter; $m(T)$ is implicitly defined by eq. (\ref{eq:self-consistency-full}).
	}
	\label{fig:plot_m_T_analytical}
\end{figure}

Consider a particle in the effective mean field potential created by other particles.
The particle can be treated as an effective two-level system, where two states correspond to sublattices $A$ and $B$.
The ground-state energy of the chosen particle is reached when it sits at the sublattice $A$:
\begin{align}
E_0 = -n_A U_A + n_B U_B
\end{align}
The particle is in the excited state, when it sits at the sublattice $B$, then its energy is
\begin{align}
E_1 = -n_B U_A + n_A U_B
\end{align}
The excitation energy of the particle is
\begin{align}
E_1-E_0 = (n_A-n_B) [U_A+U_B]=2n\, m(T) [U_A+U_B]
\end{align}
Let $n_0$ be the occupation number for the considered particle of sublattice $A$ and $n_1$ for sublattice $B$, so only one of $n_0, n_1$ is non-zero and $n_0+n_1=1$.
Now we are dealing with a simple statistical-mechanics problem of a classical two-level system with Hamiltonian $H=n_0 E_0 + n_1 E_1$, but with additional self-consistency conditions
\begin{align}
& \langle n_0 \rangle = n_A/2n \nonumber\\
& \langle n_1 \rangle = n_B/2n
\end{align}
where average is defined as $\langle O \rangle = (O_0 \cdot e^{-E_0/T} + O_1 \cdot e^{-E_1/T})/Z$, and $Z$ is the single-particle partition function $Z=e^{-E_0/T}+e^{-E_1/T}$. Subtracting these two equations we get the self-consistency condition in a compact form
\begin{align}
m(T) = 2\langle n_0 \rangle -1
\label{eq:self-consistency}
\end{align}
Substituting $\langle n_0 \rangle =  e^{-E_0/T}/Z$ to (\ref{eq:self-consistency}) we get
\begin{align}
m = \frac{2}{1+e^{-2n[U_A+U_B] m/T}} - 1
\label{eq:self-consistency-full}
\end{align}
Figure \ref{fig:plot_m_T_analytical} shows the dependence $m(T)$, implicitly defined by eq. (\ref{eq:self-consistency-full})  (this equation is used in order to construct Fig. 2a of the main text).
Expanding its RHS to the first order in $m$ we find
\begin{align}
T_{\mathrm{DW}} = \frac{N [U_A + U_B ]}{L^D}
\label{eq:T_CDW_mean_field}
\end{align}
The third-order expansion gives also the mean-field critical exponent $m(T)\sim |T-T_c|^{1/2}$).

Comparing the mean-field value of $T_{\mathrm{DW}}$ (\ref{eq:T_CDW_mean_field}) with the estimate obtained from the free energy arguments (\ref{eq:T_CDW-arbitrary-D-N}) we see that they indeed differ only by a numerical factor $2\ln2\approx 1.4$, with mean-field result giving higher value of the critical temperature.

We also note that the expression in the square brackets in (\ref{eq:T_CDW_mean_field}) is the integrated ``staggered'' strength of the interaction potential: $U_{stag} = -\sum (-1)^{i_x+i_y} U_{i_x,i_y} = U_A + U_B$. Therefore the mean-field expression for the DW-transition critical temperature can be expressed as
\begin{align}
T_{\mathrm{DW}} = \frac{N U_{stag}}{L^D}
\label{eq:T_CDW_mean_field}
\end{align}

We can estimate the limits of applicability of the mean-field theory using the Ginzburg criterion $t_G = |T_G-T_{\mathrm{DW}}|/T_{\mathrm{DW}} \sim a^2/(n \xi^2)$, where the Ginzburg temperature $T_G$ is the temperature above which fluctuation effects become important, $\xi$ is the range of the potential. The mean-field theory works better and better with the growing number of particles in $\xi \times \xi$ region, $n (\xi/a)^2$, ultimately becoming exact when the latter quantity goes to infinity.

\newpage
\section{Droplet-formation phase transition: exactly-solvable toy model and numerical tests}
\label{Appendix Collapse transition}

In this Supplementary Section we, first, consider a toy model for the droplet-formation phase transition and.

Consider $N$ particles occupying $L$ lattice sites in 1D. If two particles sit on the same lattice site they interact with energy $-U_0<0$ (on-site attraction) and particles at different sites don't interact with each other. Let $n_i$ be the occupation number of site $i$. The Hamiltonian of the system is
\begin{align}
H = -\frac{U_0}{2}\sum_{i=1}^L n_i (n_i-1) \approx -\frac{U_0}{2}\sum_{i=1}^L n_i^2
\end{align}

Such model is equivalent to the fully-connected (mean-field) Potts model with $L$ states, which possesses a phase transition from disorder (uniform occupation of all sites) to order (all particles occupy the single site): for $L=2$ it is of the second order and for $L>2$ it is of the first order \cite{Wu:1982}. The exact solution\cite{Wu:1982} gives the critical temperature:
\begin{align}
T_{\mathrm{dr}}  = \frac{L-2}{L-1} \frac{N U_0}{2\ln(L-1)}
\label{eq:Potts}
\end{align}
In case $L \gg 1$ this gives us
\begin{align}
T_{\mathrm{dr}} \approx \frac{N U_0}{2 \ln L}
\label{eq:Tc=NU0/2lnL}
\end{align}
%

The simplest model described here works for the case when on-site interaction is much greater then the interaction at nearest-neighboring sites, which corresponds to the case of infinitely many degenerate modes in the optical cavity. Alternatively it can be viewed as a coarse-grained model for interaction with range $\xi$, where one site corresponds a coarse-grained cell of the size $\xi$, so we should substitute $L\rightarrow L/\xi$. Additional substitution $L/\xi \rightarrow (L/\xi)^D$ trivially generalizes eq. (\ref{eq:Tc=NU0/2lnL}) to the $D$-dimensional case, thus we arrive to the formula (2) of the main text.

\begin{figure}[!b]
	\centering
	\subfloat[]{\includegraphics[width=0.45\linewidth]{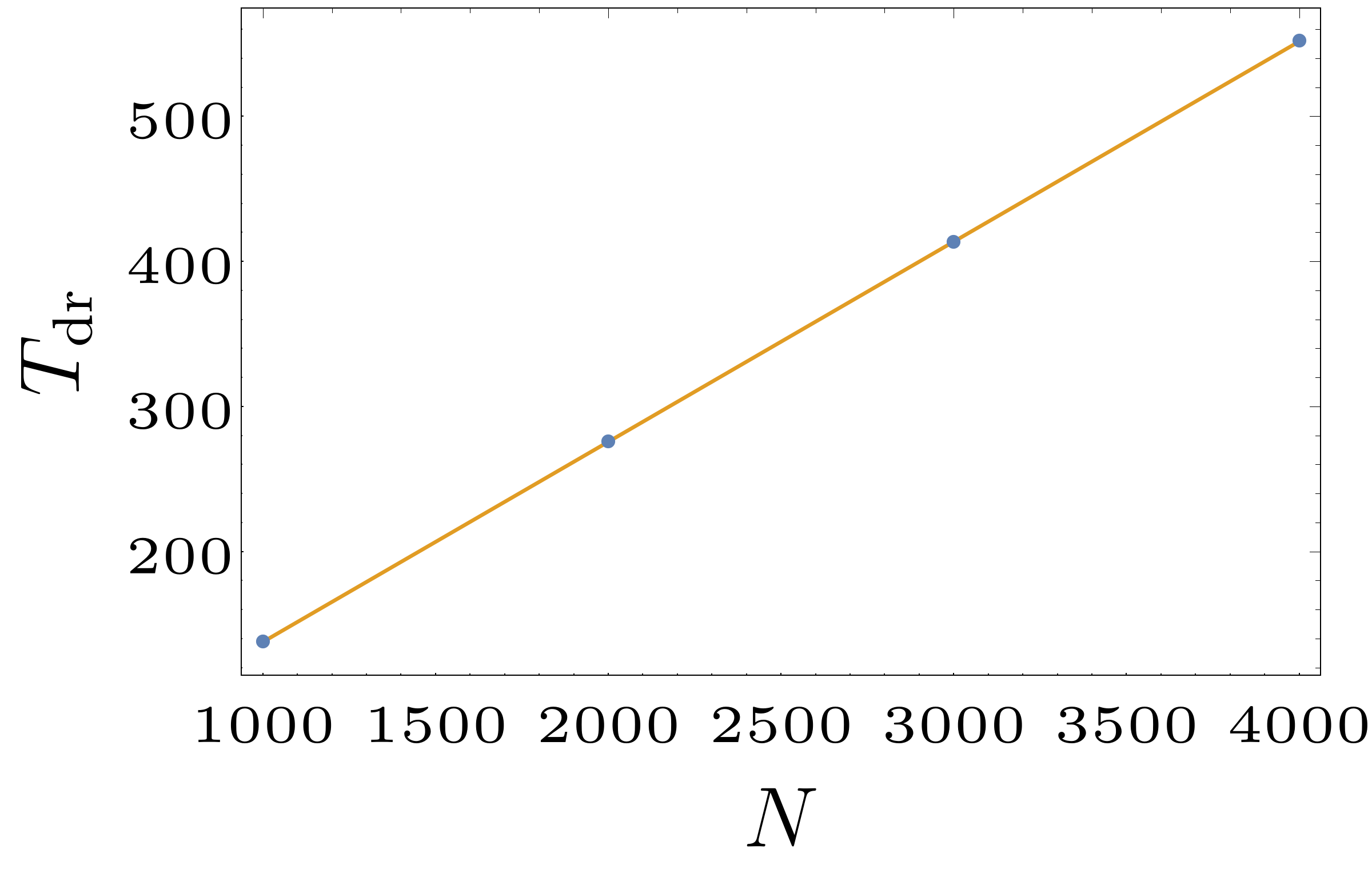}}
	\qquad
	\subfloat[]{\includegraphics[width=0.45\linewidth]{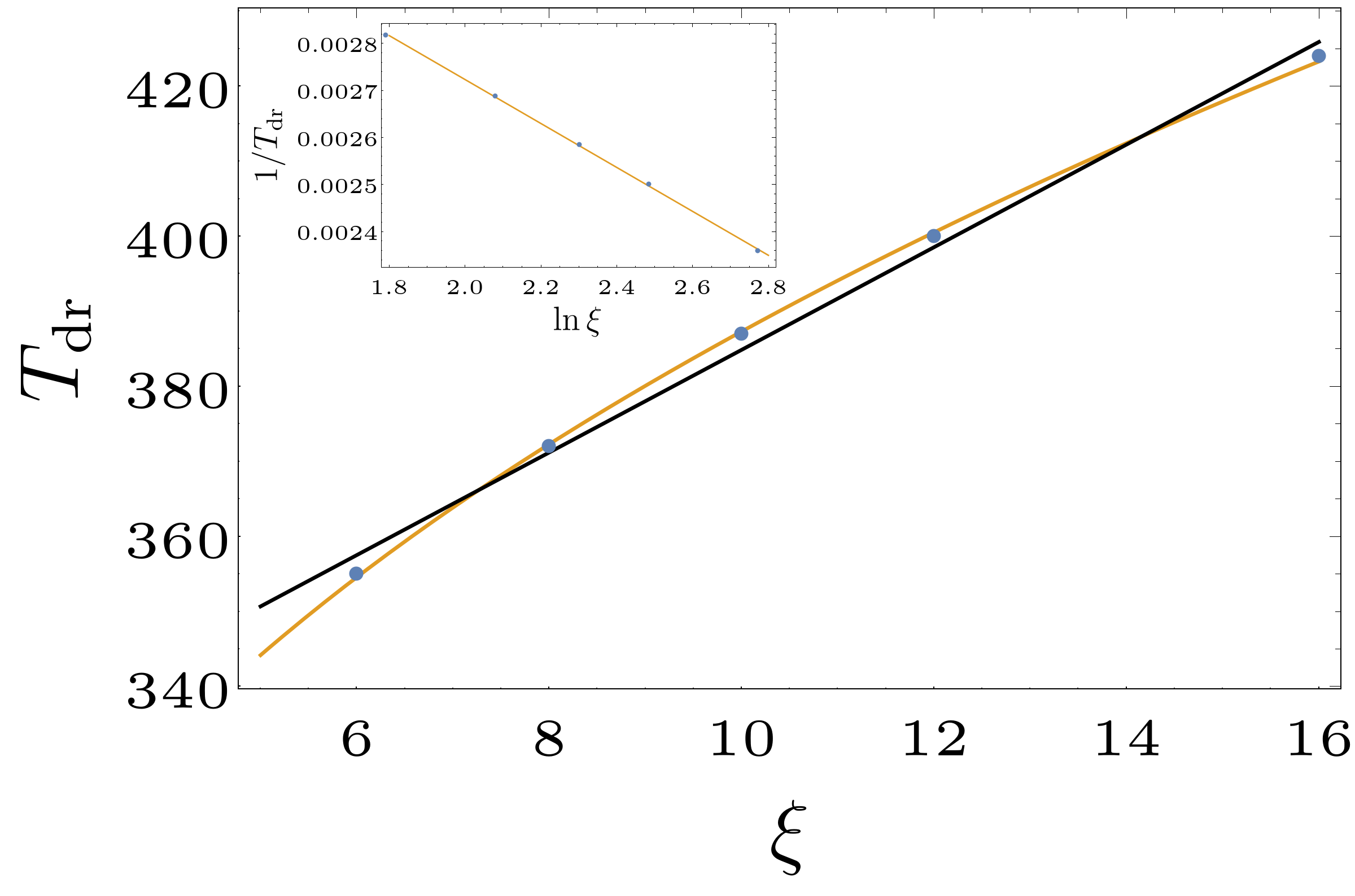}}
	\caption{Dependence of the droplet-transition temperature on $T_{\mathrm{dr}}$ on a) the number of particles $N$; b) the interaction range $\xi$.
		a) Blue dots are the data points obtained by MC simulations, orange line is the linear fit. Simulations were performed for 1D system with $L=100$, $\xi=5$.
		b) Orange curve shows fit according to eq. $T_{\mathrm{dr}}=A/\ln(B/\xi)$, black line
		shows for comparison the best linear fit $T_{\mathrm{dr}}=m \xi+b$. While it is hard to distinguish log vs linear fits, the
		concave character of the function can be clearly seen. MC simulations were performed for 1D system of
		size $L=300$, number of particles $N=3000$.
		Here $T_{\mathrm{dr}}$ is measured in units of $U_0$; $\xi$ and $L$  are measured in the units of $a$.
	Standard deviations of the measurements are smaller than the size of the symbols.}
	\label{fig:plots_Tdr}
\end{figure}

In order to test eq. (2) of the main text we perform Monte Carlo (MC) simulations of the 1D system with the same Gaussian potential with the range $\xi$ as we use in the main text. Since eq. (2) works only qualitatively, we check the linear dependence of $T_{\mathrm{dr}}$ on the number of particles $N$  (Fig. \ref{fig:plots_Tdr}a) and the dependence of the form $T_{\mathrm{dr}} = A / \ln(B/\xi)$ (Fig. \ref{fig:plots_Tdr}b).

\newpage

\section{Simulation of interaction of color-charged droplets}
\label{Appendix Droplets interaction}

\begin{figure}[tbh]
	\centering
	\subfloat[initial configuration]{\includegraphics[width=0.256\linewidth]{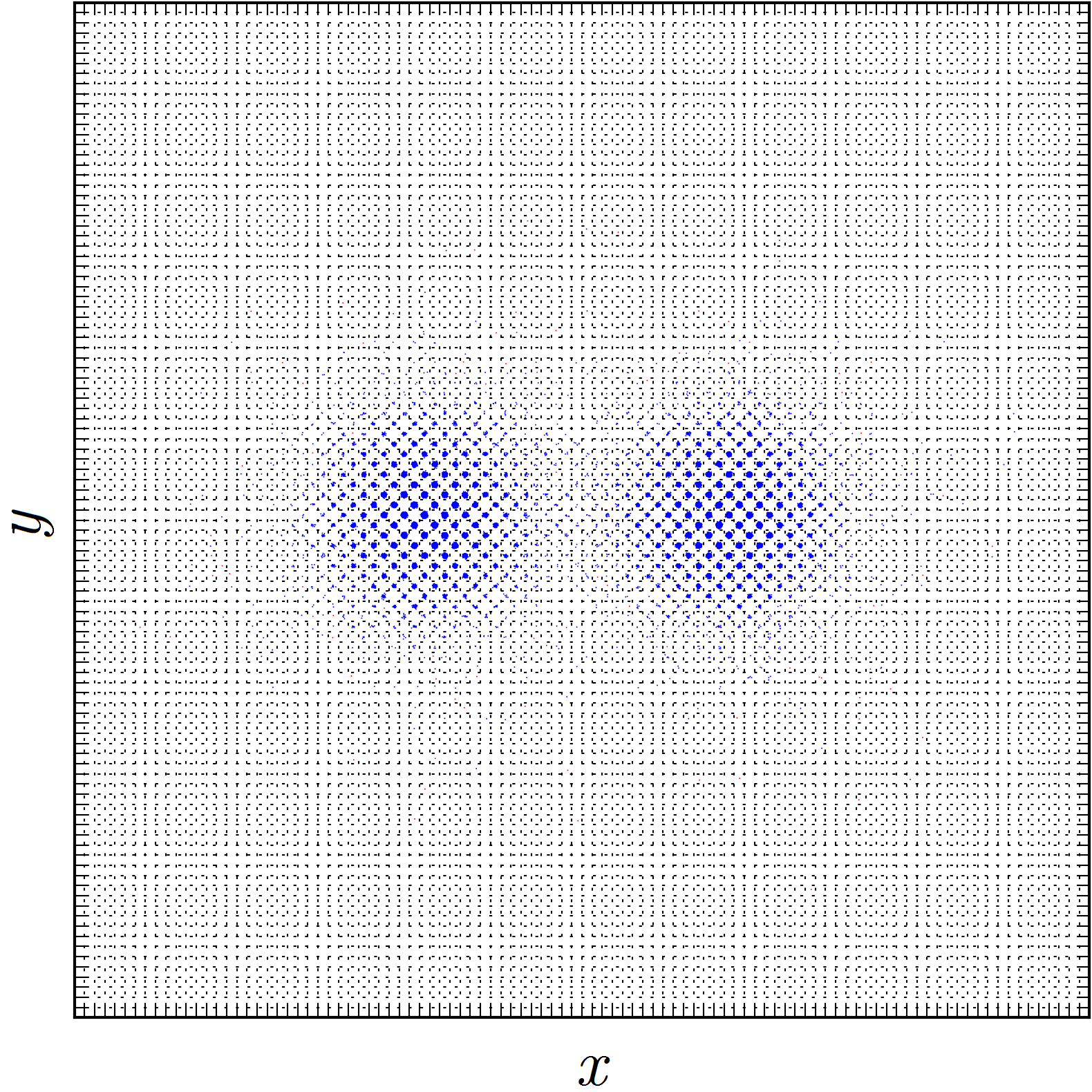}}
	\hfill
	\subfloat[after 500 MC sweeps]{\includegraphics[width=0.24\linewidth]{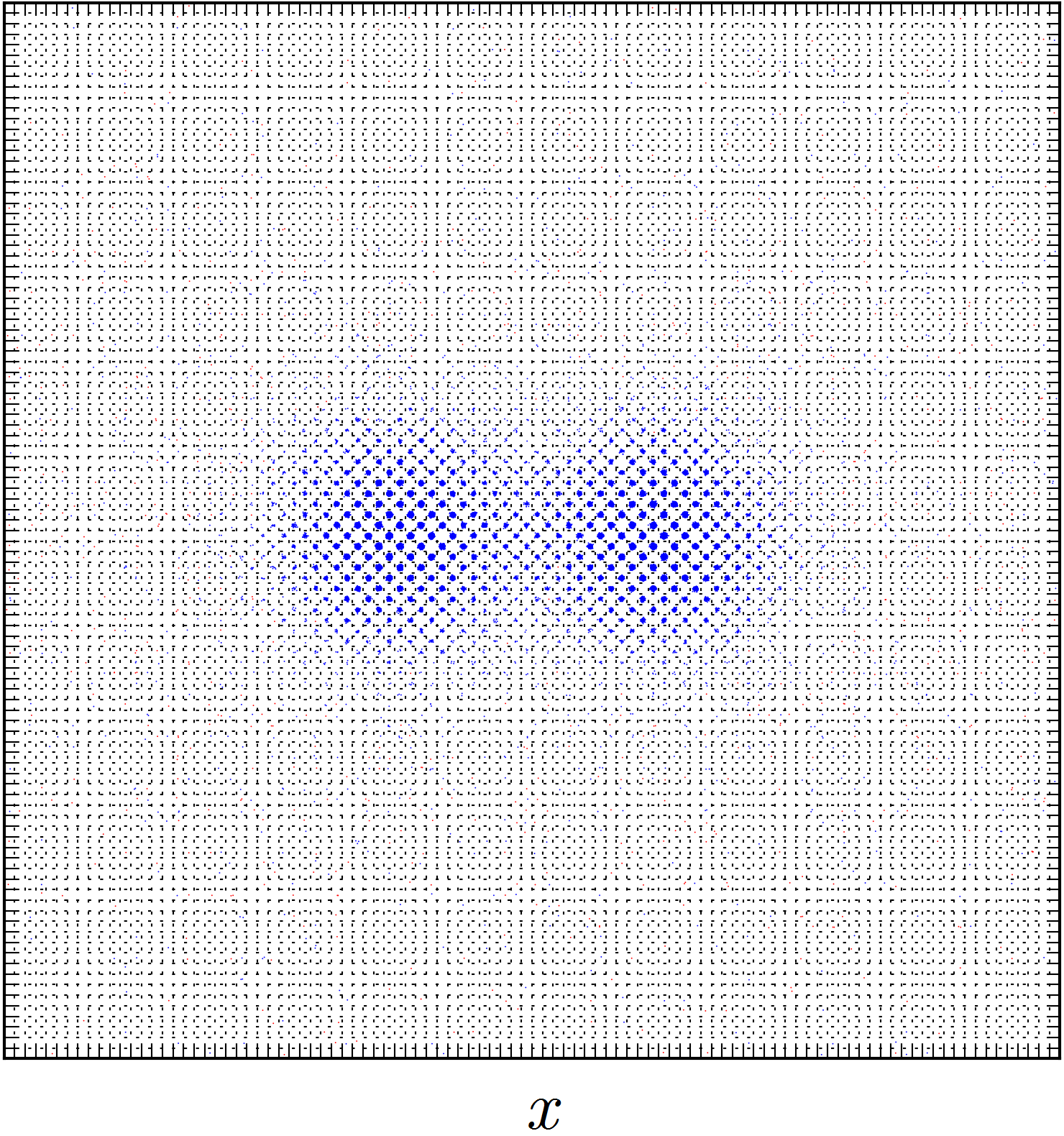}}
	\hfill
	\subfloat[after 600 MC sweeps]{\includegraphics[width=0.24\linewidth]{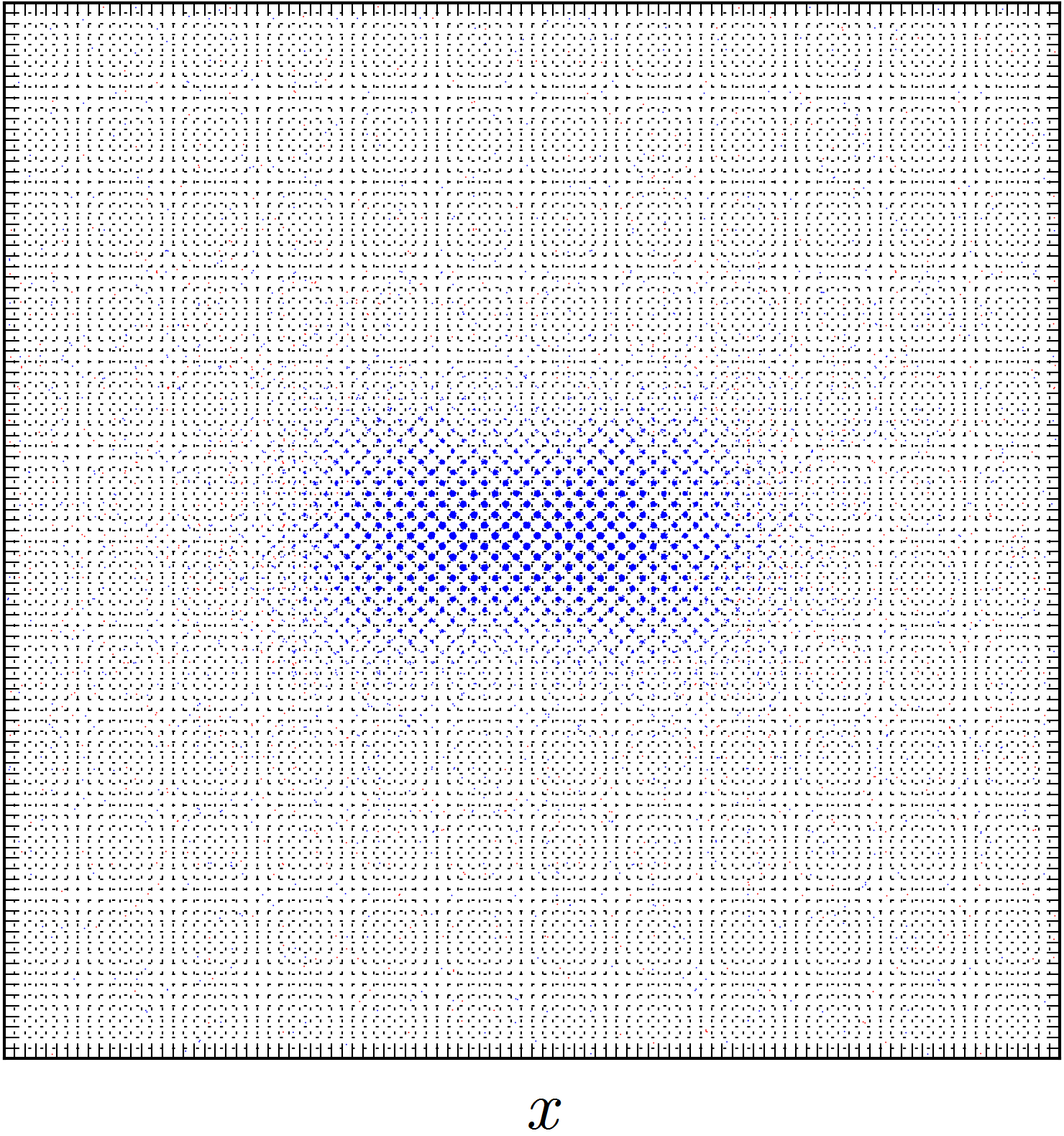}}
	\hfill
	\subfloat[after 1000 MC sweeps]{\includegraphics[width=0.24\linewidth]{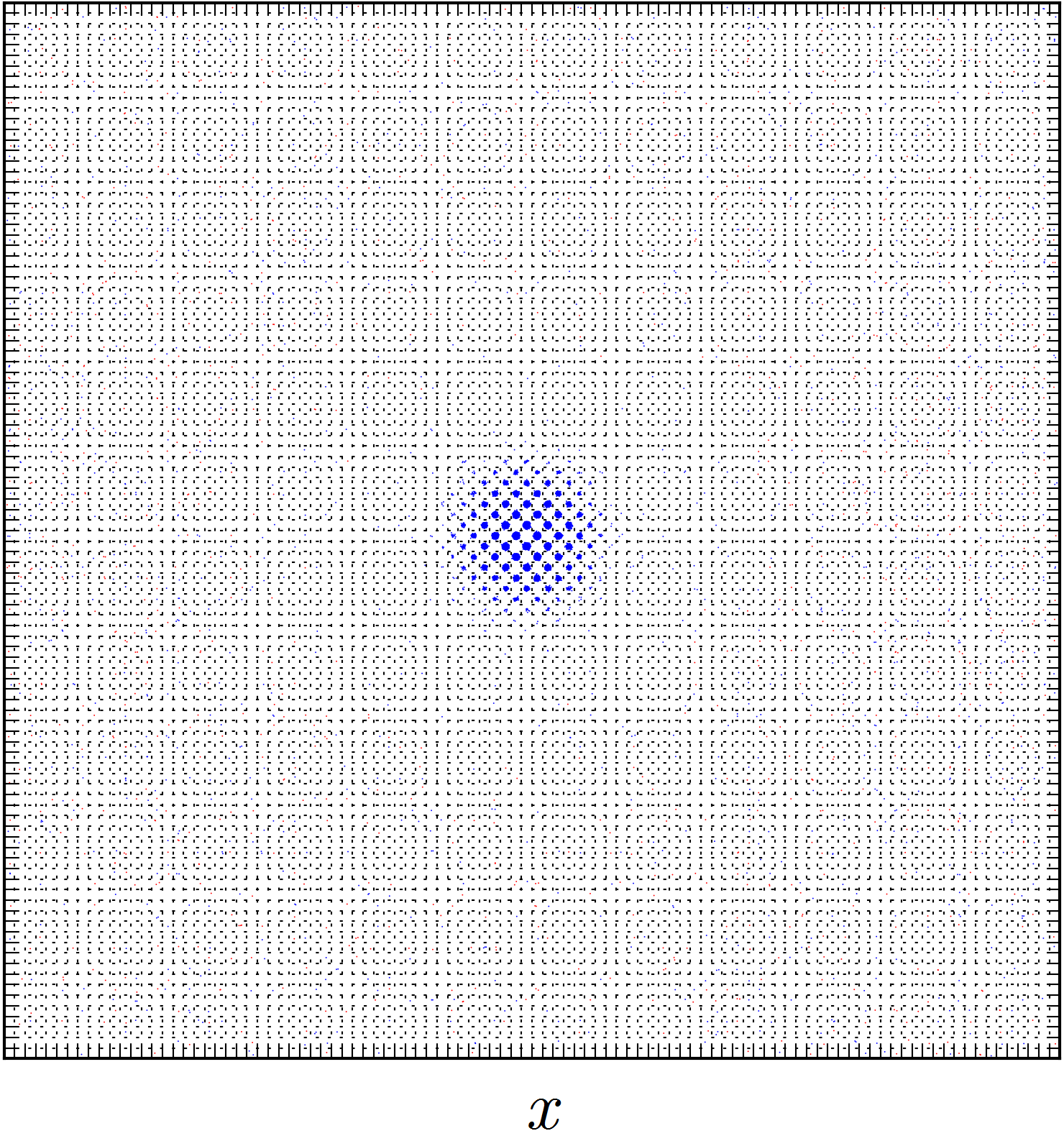}}
	\hfill
	\subfloat[initial configuration]{\includegraphics[width=0.256\linewidth]{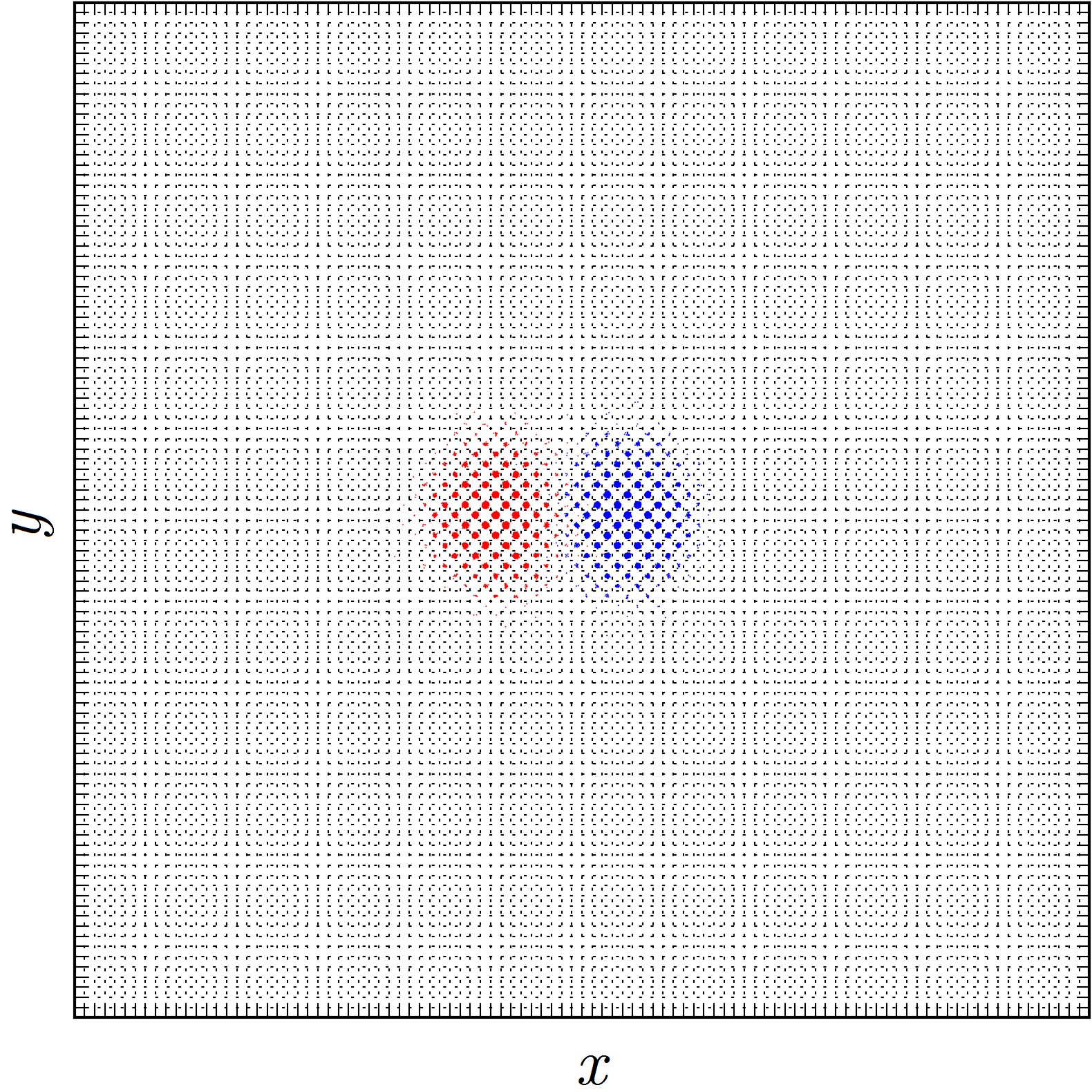}}
	\quad
	\subfloat[after 100 MC sweeps]{\includegraphics[width=0.24\linewidth]{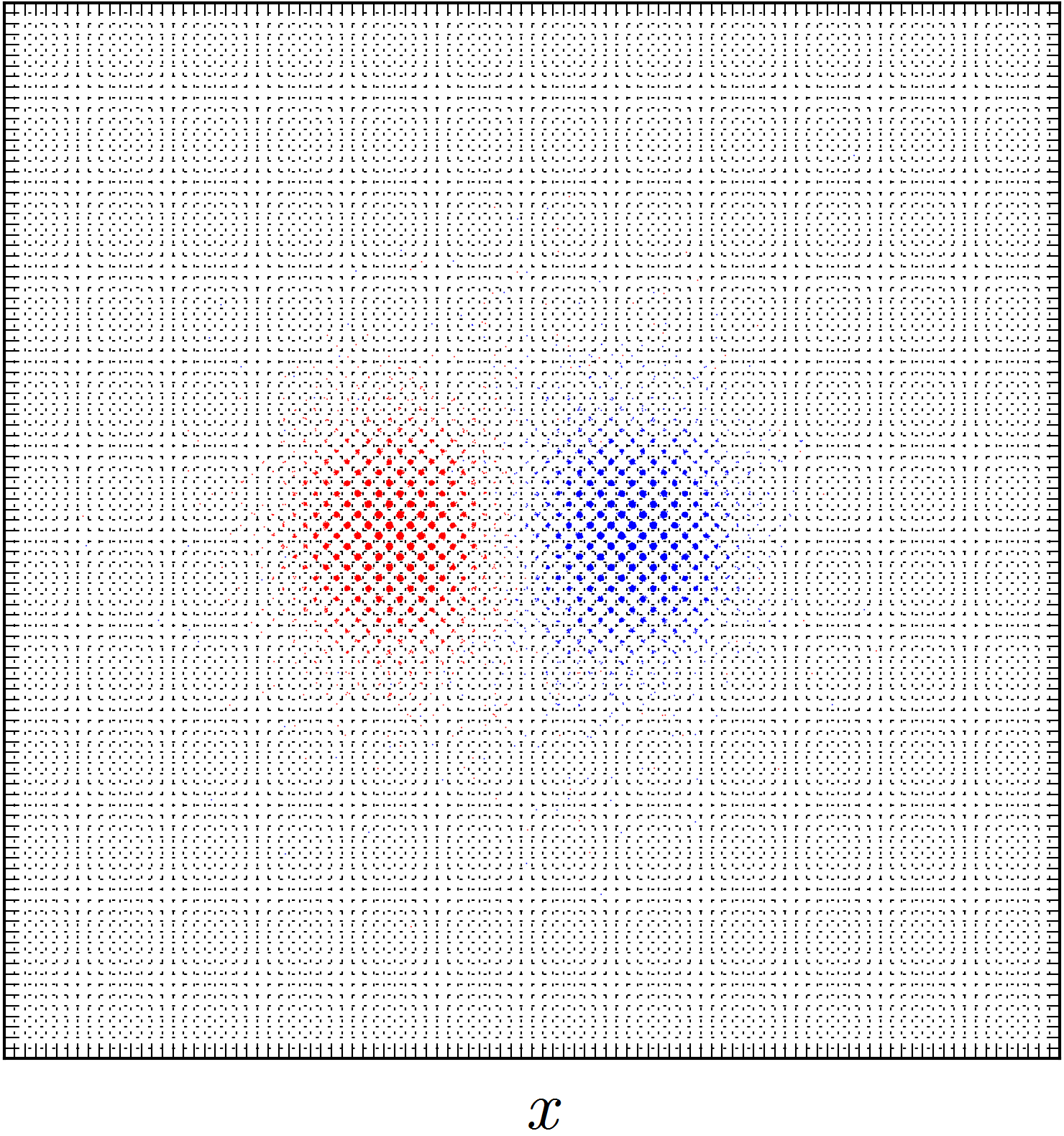}}
	\quad
	\subfloat[after 1000 MC sweeps]{\includegraphics[width=0.24\linewidth]{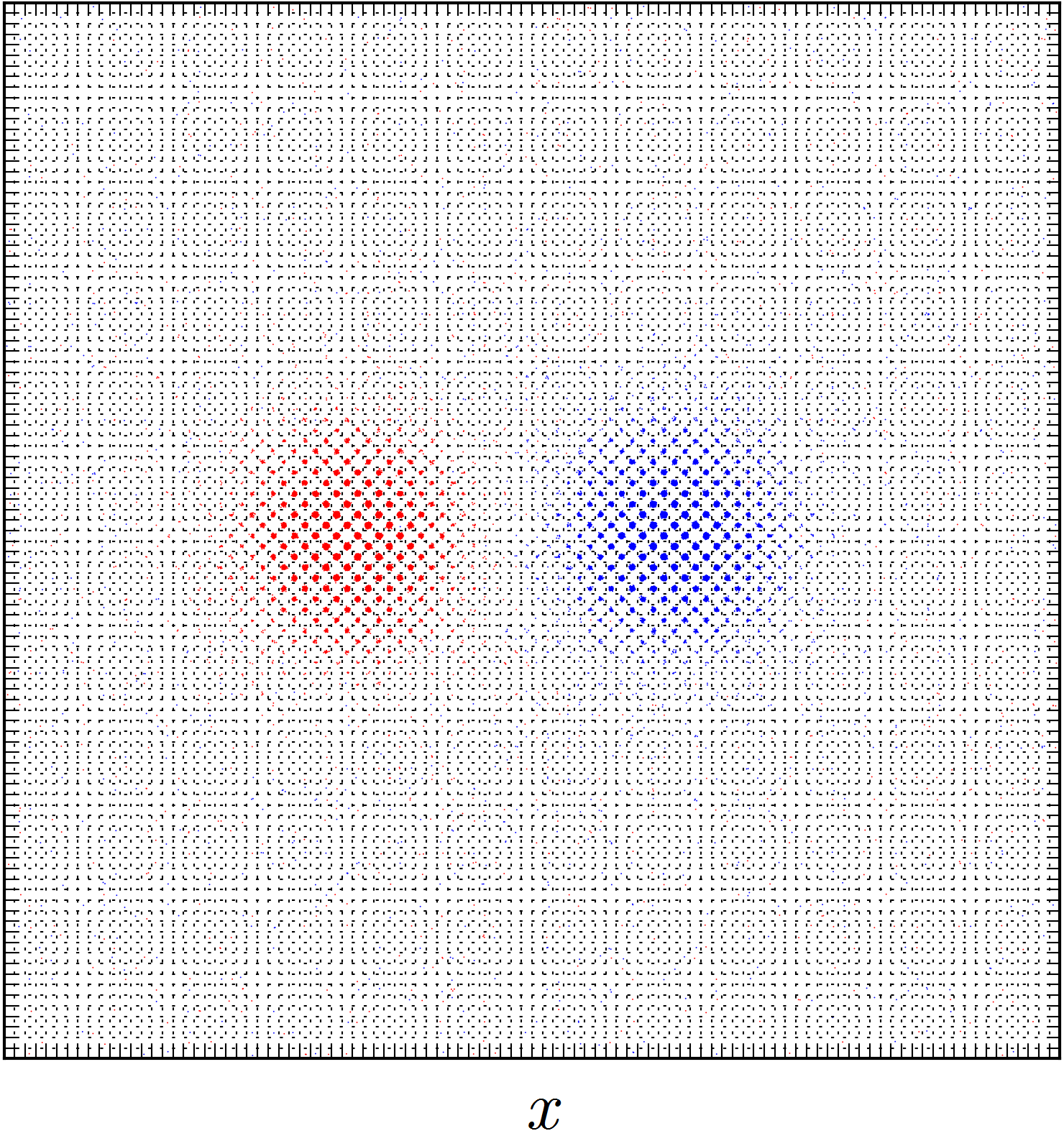}}
	\caption{Interaction of droplets: attraction of equally charged droplets (upper panel) and repulsion of oppositely charged ones (lower panel). For each case MC evolution at a constant temperature is shown. Parameters of the simulation: $L=100$, $\xi=10$.}
	\label{fig:attraction,repulsion}
\end{figure}

\section{Derivation of the light-mediated interaction of atoms and mapping to the classical lattice gas problem}
\label{Appendix Derivation of the model}

\begin{figure}[tbh] %
	\centering
	\centering
	\includegraphics[width=0.5\linewidth]{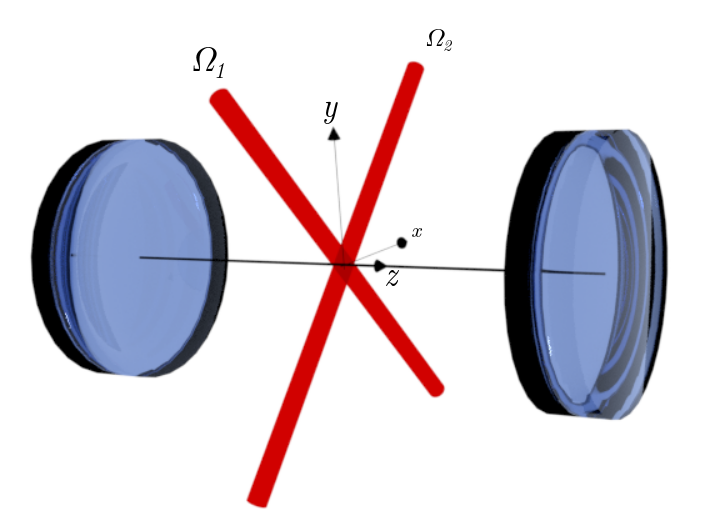}
	\caption{Geometry of the system. Cavity axis is parallel to $z$; both lasers and the atomic gas are in the $xy$ plane.
	}
	\label{fig:geometry}
\end{figure}

In this Supplementary Section we describe the setting we propose
for the experimental realization of the interaction we use in the main
text. We consider an ensemble of atoms in a multimode cavity pumped transverse to the cavity axis by
a pair of lasers with a frequency $\omega_L$ and mode functions
$\mathbf{\Omega}(\mathbf{r}) = \mathbf{e}_z \Omega(\mathbf{r}) \equiv
\mathbf{e}_z (\Omega_1(\tilde{x}) + \Omega_2(\tilde{y})) \equiv  \mathbf{e}_z
(\cos(2\pi \tilde{x}/\lambda) + \cos(2\pi \tilde{y}/\lambda))$ (here $\tilde{x}=(x-y)/\sqrt{2}$, $\tilde{y}=(x+y)/\sqrt{2}$, Fig. \ref{fig:geometry}).
The cavity possesses (nearly) degenerate modes with frequencies $\omega_{c\alpha}$ (where $\alpha$ is the mode index).
The atoms are treated as two-level systems with the level splitting $\omega_A$ and the transition dipole moment connecting the ground and excited state being parallel to $\mathbf{e}_x+\mathbf{e}_z$ (spatial isotropy can be broken by external magnetic field), which allows to couple the atomic dipole transitions to both pump and cavity modes. For more realistic experimental settings with many-level atoms such coupling can be achieved by employing vector polarizability of the atoms\cite{LeKien:2013}.

For low density of atoms we can neglect their contact interaction and
describe the light-mediated interaction in a multimode cavity by the following effective action \cite{Gopalakrishnan:2009}:
\begin{align}
S = S_{at} + S_{int} + S_{em}
\end{align}
\begin{align}
S_{at} =  \int d\mathbf{r} d\tau \left[ \Psi_g^*(\mathbf{r},\tau) \left( \partial_{\tau} -\frac{\hbar\nabla^2}{2M} -\frac{\mu}{\hbar} \right) \Psi_g(\mathbf{r},\tau) +
	\Psi_e^*(\mathbf{r},\tau) \left( \partial_{\tau} -\frac{\hbar\nabla^2}{2M} -\frac{\mu}{\hbar} + \Delta_a \right) \Psi_e(\mathbf{r},\tau)  \right]
\end{align}
\begin{align}
S_{int} = \int d\mathbf{r} d\tau \left[ \sum_{\alpha} i g_{\alpha}(\mathbf{r}) \Psi^*_e(\mathbf{r},\tau) \Psi_g(\mathbf{r},\tau) a_{\alpha}(\tau) + i\Omega(\mathbf{r}) \Psi^*_e (\mathbf{r},\tau) \Psi_g(\mathbf{r},\tau) + h.c.\right]
\end{align}
\begin{align}
S_{em} = \int d\tau \sum_{\alpha} a^*_{\alpha}(\tau) (\partial_{\tau}+\Delta_{c\alpha}) a_{\alpha}(\tau)
\end{align}
Here $\Psi_g$ and $\Psi_e$ are the bosonic second-quantized fields corresponding to the ground and excited states of the atoms with mass $M$. We work in the reference frame rotating at the laser frequency $\omega_L$ and denote $\Delta_a = \omega_A - \omega_L$, $\Delta_{c\alpha} = \omega_{c\alpha} - \omega_L$. $g_{\alpha} (\mathbf{r}) = g_{0} \Xi_{\alpha}(\mathbf{r})$ where $g_0$ is the atom-cavity coupling strength and $\Xi_{\alpha}(\mathbf{r})$ is the normalized mode function of the cavity mode $\alpha$.
Integrating out excited state $\Psi_e$ and cavity modes $a_{\alpha}$ we get the following effective action
\begin{align}
S_{\text{eff}}[\Psi_g,\Psi_g^*]&=\int d\mathbf{r} d\tau \Psi_g^*(\mathbf{r},\tau) \left( \partial_{\tau} -\frac{\nabla^2}{2M} +(V(\mathbf{r})-\mu) \right ) \Psi_g(\mathbf{r},\tau)-\nonumber \\
&+ \int d\tau d\mathbf{r} d\mathbf{r}'
|\Psi_g(\mathbf{r},\tau)|^2 U(\mathbf{r},\mathbf{r}')  |\Psi_g(\mathbf{r}',\tau)|^2
\label{eq:eff-action-simplified}
\end{align}
The external optical potential $V(\mathbf{r})$ and the cavity-mediated interaction $U(\mathbf{r}, \mathbf{r}')$ are given by
\begin{align}
V(\mathbf{r}) &= -\frac{1}{\Delta_a} |\Omega(\mathbf{r})|^2 \label{eq:ExternalPotential}\\
U(\mathbf{r},\mathbf{r}') &=- \frac{g_0^2}{\Delta_a^2} \Omega^*(\mathbf{r}) \Omega(\mathbf{r}') \sum_{\alpha}  \frac{\Xi^*_{\alpha}(\mathbf{r}) \Xi_{\alpha}(\mathbf{r}')}{\Delta_{c\alpha}} \label{eq:InteractionPotential}
\end{align}
We consider the case of red-detuned laser frequency with respect to both atom splitting and to cavity nearly degenerate modes $\Delta_a, \Delta_{c\alpha}>0$. In this case the atoms are attracted to maxima of $|\Omega(\mathbf{r})|^2$; the interaction is attractive at small distances $\mathbf{r}'\rightarrow \mathbf{r}$ i.e. $U(\mathbf{r},\mathbf{r})<0$, which creates a possibility for a peculiar droplet-formation phase transition.

{\it Mapping to a classical lattice-gas problem.} We consider the
classical limit of the action (\ref{eq:eff-action-simplified}), which
bounds of applicability we find below. For a large number of particles $N\gg1$ we can use the canonical ensemble ($N=const$) instead of the grand canonical one ($\mu=const$), and describe the system of atoms by a classical Hamiltonian
$H = \sum_i V(\mathbf{r}_i) + \sum_{i ,j} U(\mathbf{r}_i, \mathbf{r}_j)$.%

We study the case when the external optical potential $V$ confines all the particles to the plane $z=0$.
For the pump mode function chosen as $\Omega(\mathbf{x}) = \Omega_0 (\cos(2\pi \tilde{x}/\lambda)+ \cos(2\pi \tilde{y}/\lambda))$,
the external optical potential (\ref{eq:ExternalPotential}) becomes
$V(\mathbf{x})= -V_0 (\cos(2\pi \tilde{x}/\lambda)+\cos (2\pi \tilde{y}/\lambda))^2$, with $V_0 = \Omega_0^2/\Delta_a^2$.
The interaction potential (\ref{eq:InteractionPotential})  takes the form $U(\mathbf{x},\mathbf{x}') \sim - (\cos\frac{2\pi \tilde{x}}{\lambda}+\cos\frac{2\pi \tilde{y}}{\lambda}) (\cos\frac{2\pi \tilde{x}'}{\lambda}+\cos\frac{2\pi \tilde{y}'}{\lambda})$, see Fig. 1.
The interaction potential $U(\mathbf{x}, \mathbf{x}')$ is
sign-alternating and attractive at short distances, as considered
  in the main text.
For $V_0 \gg|U|$ it is convenient to define the lattice version of the interaction
potential $U_{i, j}=U(0,i \,\mathbf{a}_1+j  \,\mathbf{a}_2)$ where
$\mathbf{a}_1, \mathbf{a}_2$ are the basis vectors of the optical
lattice, with $|\mathbf{a}_1|=|\mathbf{a}_2|=a=\lambda/\sqrt{2}$ (Fig. 1a).

\section{Experimental realization and results for the 4-color case.}
\label{Appendix Z4}

In this Supplementary Section we show how to modify the experimental scheme in order to have 4 types of color-charges and present the results of the simulations in this case.

\begin{figure}[tbh] %
	\centering
	\includegraphics[width=5cm]{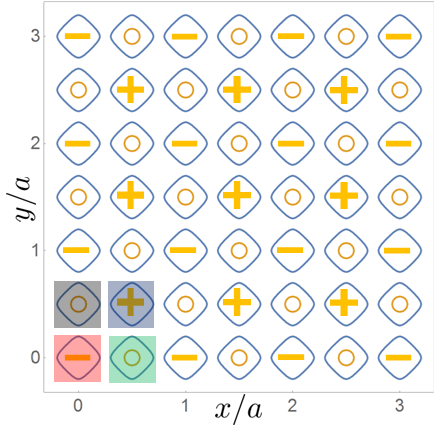}
	\caption{Contour plot of the external potential $V(x,y)$ (blue contours) and signs of the interaction potential $U(0,0,x,y)$ ($+,-,0$) for the 4-color  case. The contours encircle the local minima of the potential $V$. The interaction potential $U$ favors the formation of density-wave corresponding to one of the four sublattices $A$, $B$, $C$, $D$ (marked by red, green, black, and blue colors) and endows the droplets with $\mathbb{Z}_2 \times \mathbb{Z}_2$ color-charge.}
	\label{fig:potential_orthogonal_polarizations}
\end{figure}

We consider a pair of transverse to the cavity axis lasers with orthogonally-polarized pump modes $\mathbf{\Omega}(\mathbf{r}) =  (\mathbf{e}_y\Omega_1(x) + \mathbf{e}_x\Omega_2(y)) \equiv  (\mathbf{e}_y\cos(2\pi x/\lambda) + \mathbf{e}_x\cos(2\pi y/\lambda))$, with each polarization being coupled to one of the two degenerate excited states $\Psi_{e,1}$, $\Psi_{e,2}$, which scatter the light into two families of orthogonally-polarized cavity modes $a_{\alpha,1}$, $a_{\alpha,2}$ degenerate with respect to polarization.
Analogously to the previous section we have

\begin{align}
S_{at} =  \int d\mathbf{r} d\tau \left[ \Psi_g^*(\mathbf{r},\tau) \left( \partial_{\tau} -\frac{\hbar\nabla^2}{2M} -\frac{\mu}{\hbar} \right) \Psi_g(\mathbf{r},\tau) +
	\sum_{i=1,2}  \Psi_{e,i}^*(\mathbf{r},\tau) \left( \partial_{\tau} -\frac{\hbar\nabla^2}{2M} -\frac{\mu}{\hbar} + \Delta_a \right) \Psi_{e,i}(\mathbf{r},\tau)  \right]
\end{align}
\begin{align}
S_{int} = \int d\mathbf{r} d\tau \left[ i \sum_{\alpha,i} g_{\alpha}(\mathbf{r}) \Psi^*_{e,i}(\mathbf{r},\tau) \Psi_g(\mathbf{r},\tau) a_{\alpha,i}(\tau) + i\sum_{i=1,2}\Omega_i(\mathbf{r}) \Psi^*_{e,i} (\mathbf{r},\tau) \Psi_g(\mathbf{r},\tau) + h.c.\right]
\end{align}
\begin{align}
S_{em} = \int d\tau \sum_{\alpha,i} a^*_{\alpha,i}(\tau) (\partial_{\tau}+\Delta_{c\alpha}) a_{\alpha,i}(\tau)
\end{align}
After integrating out excited state $\Psi_e$ and cavity modes $a_{\alpha}$ we get the effective action (\ref{eq:eff-action-simplified})
with the external and interaction potentials
\begin{align}
V(\mathbf{r}) &= -\frac{1}{\Delta_a} \left(|\Omega_1(\mathbf{r})|^2 +|\Omega_2(\mathbf{r})|^2 \right) \label{eq:ExternalPotential}\\
U(\mathbf{r},\mathbf{r}') &= -\frac{g_0^2}{\Delta_a^2} \left(\Omega_1^*(\mathbf{r}) \Omega_1(\mathbf{r}') + \Omega_2^*(\mathbf{r}) \Omega_2(\mathbf{r}')\right)\sum_{\alpha}  \frac{\Xi^*_{\alpha}(\mathbf{r}) \Xi_{\alpha}(\mathbf{r}')}{\Delta_{c\alpha}} \label{eq:InteractionPotential}
\end{align}
Periodicity of the external potential $V$ and the sign-changing nature of the interaction $U$ is shown in Fig. \ref{fig:potential_orthogonal_polarizations}. DW-state here breaks the translational  $\mathbb{Z}_2 \times \mathbb{Z}_2$ symmetry.

\begin{figure}[tbh]
	\centering
	\subfloat[]{\includegraphics[height=4cm]{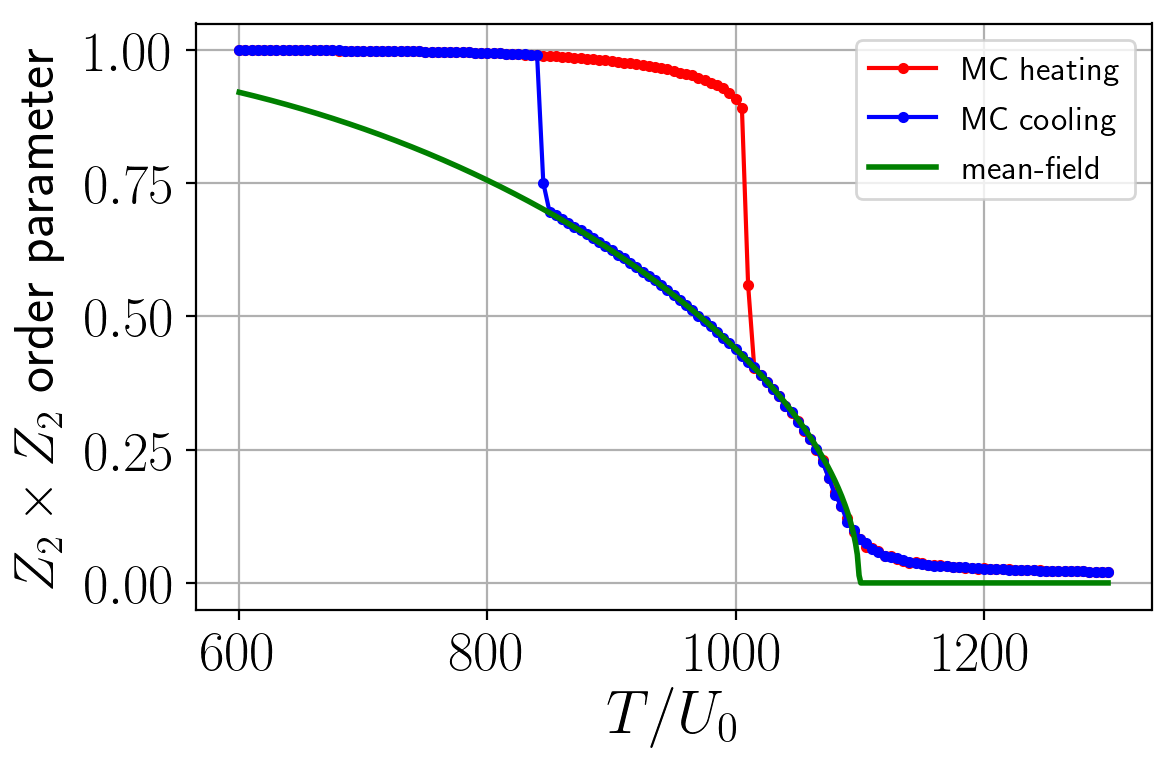}}
	\qquad
	\subfloat[]{\includegraphics[height=4cm]{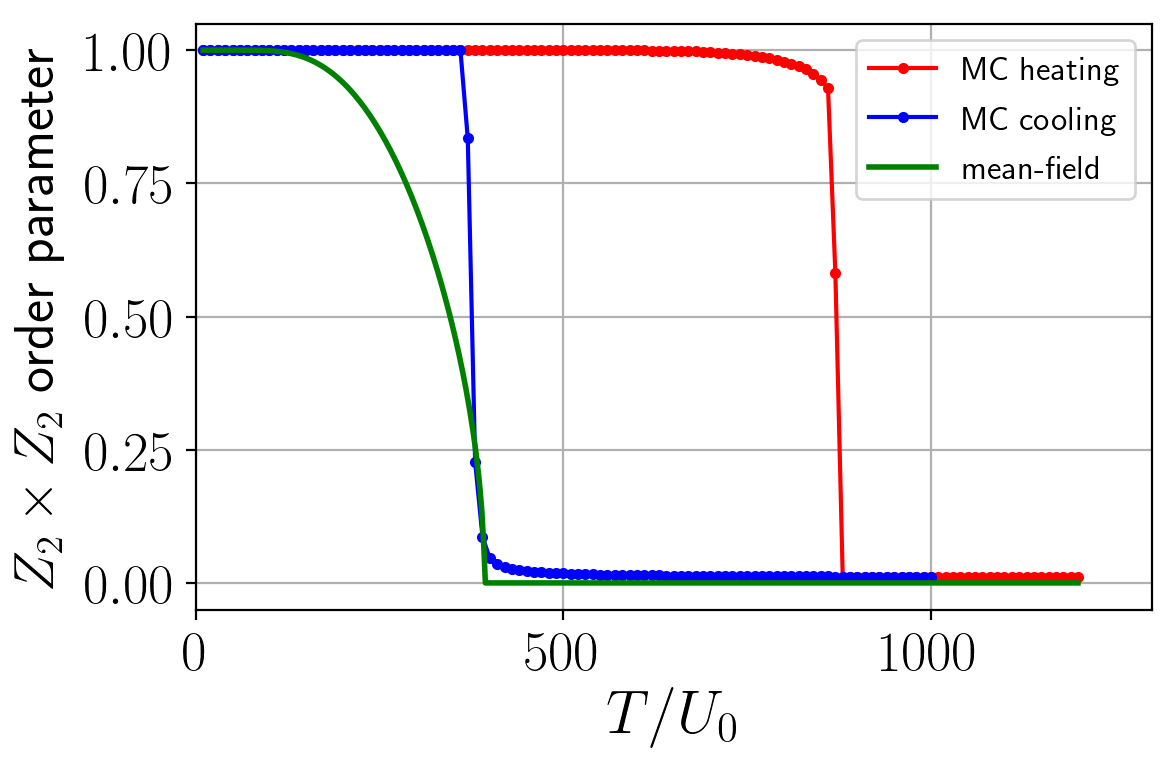}}
	\\
	\subfloat[]{\includegraphics[height=4cm]{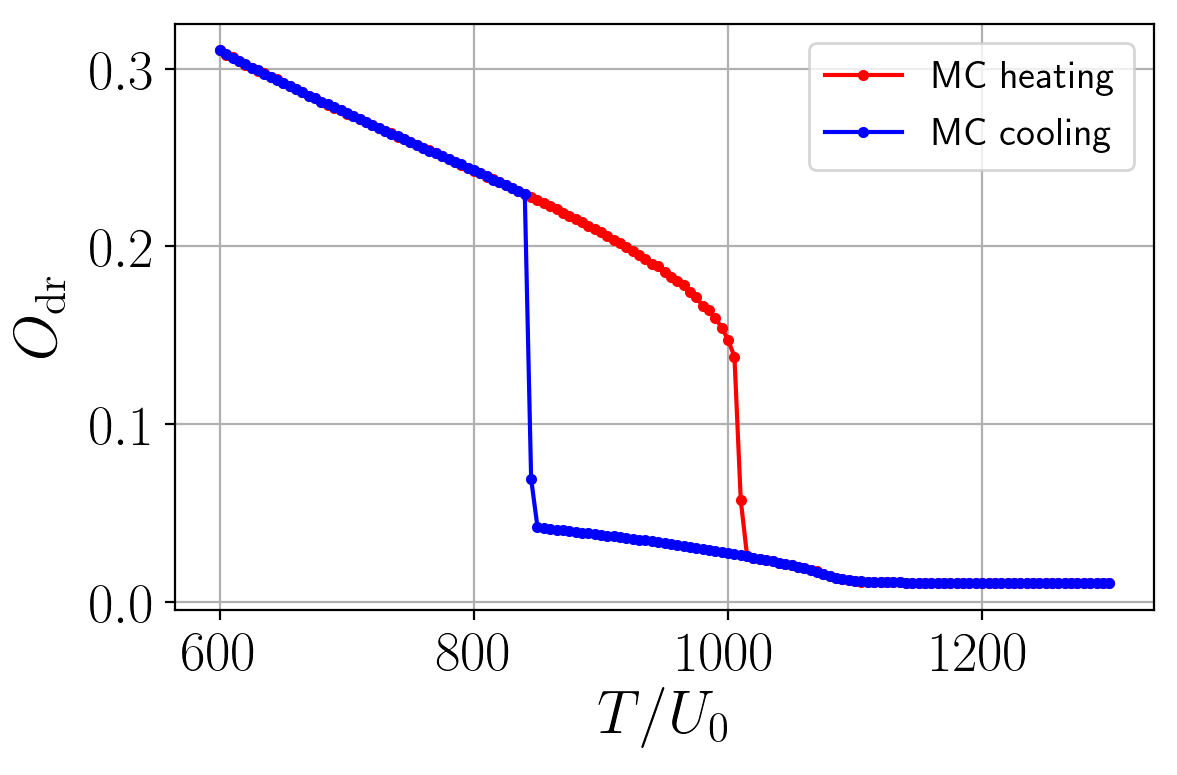}}
	\qquad
	\subfloat[]{\includegraphics[height=4cm]{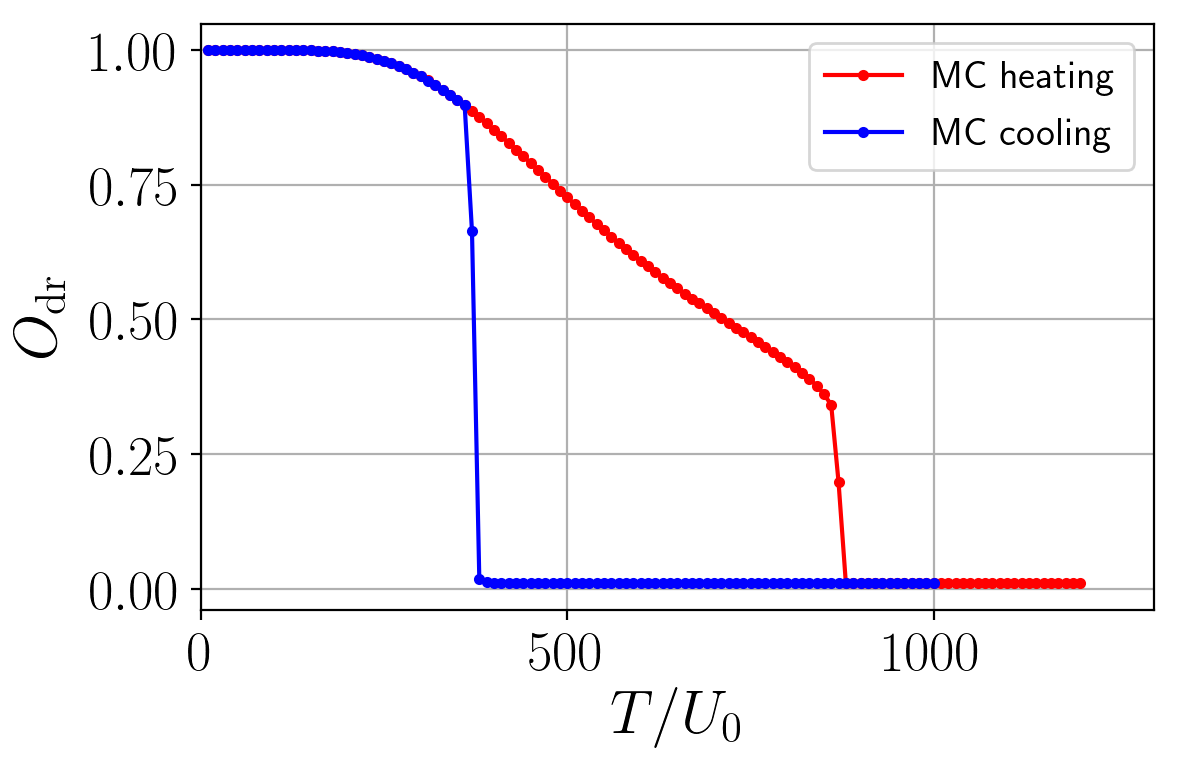}}
	\caption{Temperature dependence of the order parameters for the 4-color case for $\xi=8.5$ (left panel) and $\xi=5$ (right panel); $L=30$. (a,b) $\mathbb{Z}_2 \times \mathbb{Z}_2$-sublattice imbalance order parameter: $\sqrt{\frac{4}{3}\sum_{\alpha}(\frac{N_{\alpha}}{N}-\frac{1}{4})^2}$, where $N_{\alpha}$ is the number of particles occupying sublattice $\alpha=A,B,C,$ or $D$. (c,d) Droplet order parameter.
		}
	\label{fig:Z4_modeling_plots}
\end{figure}

\begin{figure}[!t]
	\centering
	\subfloat[uniform]{\includegraphics[width=0.348\linewidth]{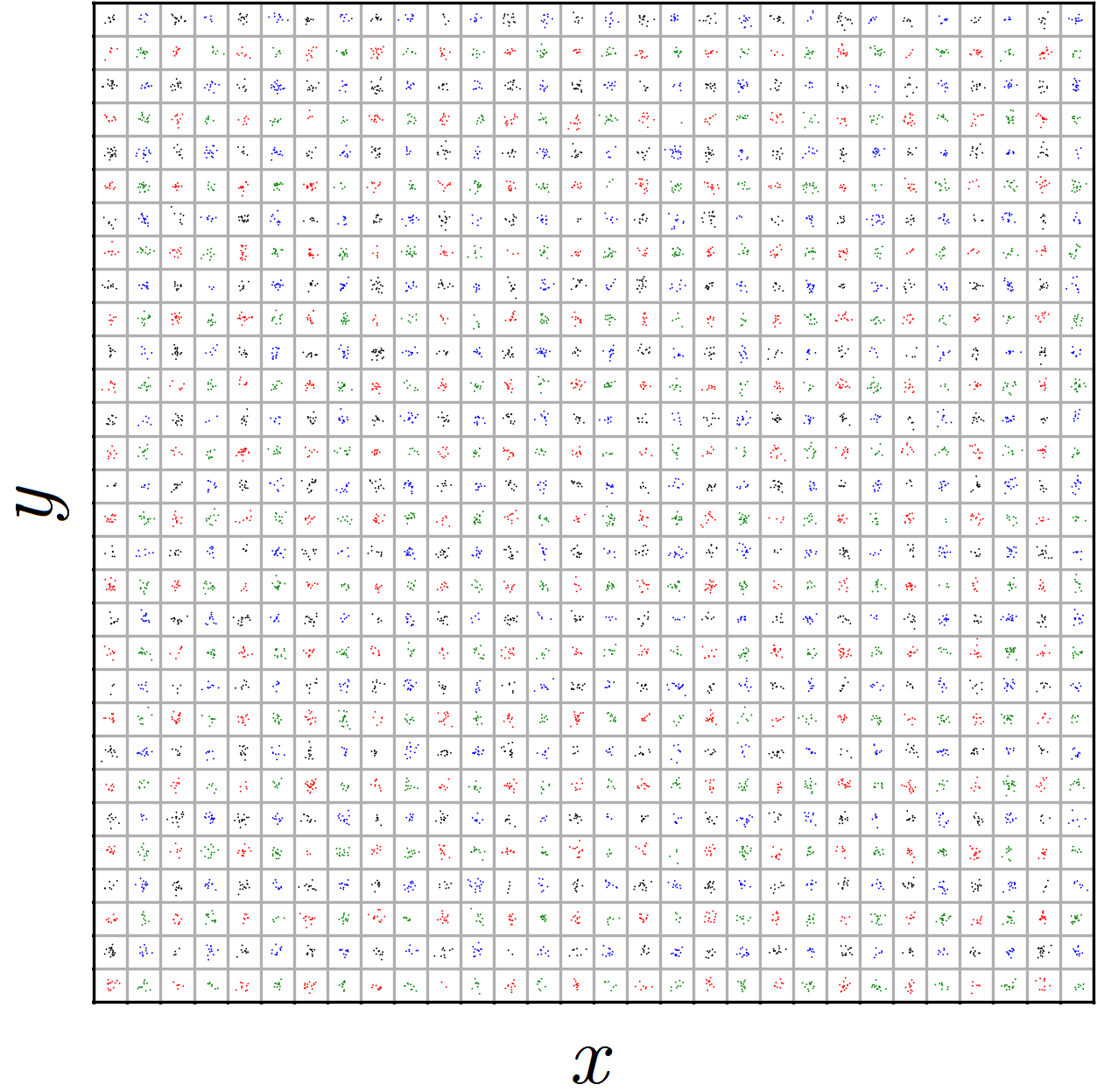}}
	\hfill
	\subfloat[DW]{\includegraphics[width=0.32\linewidth]{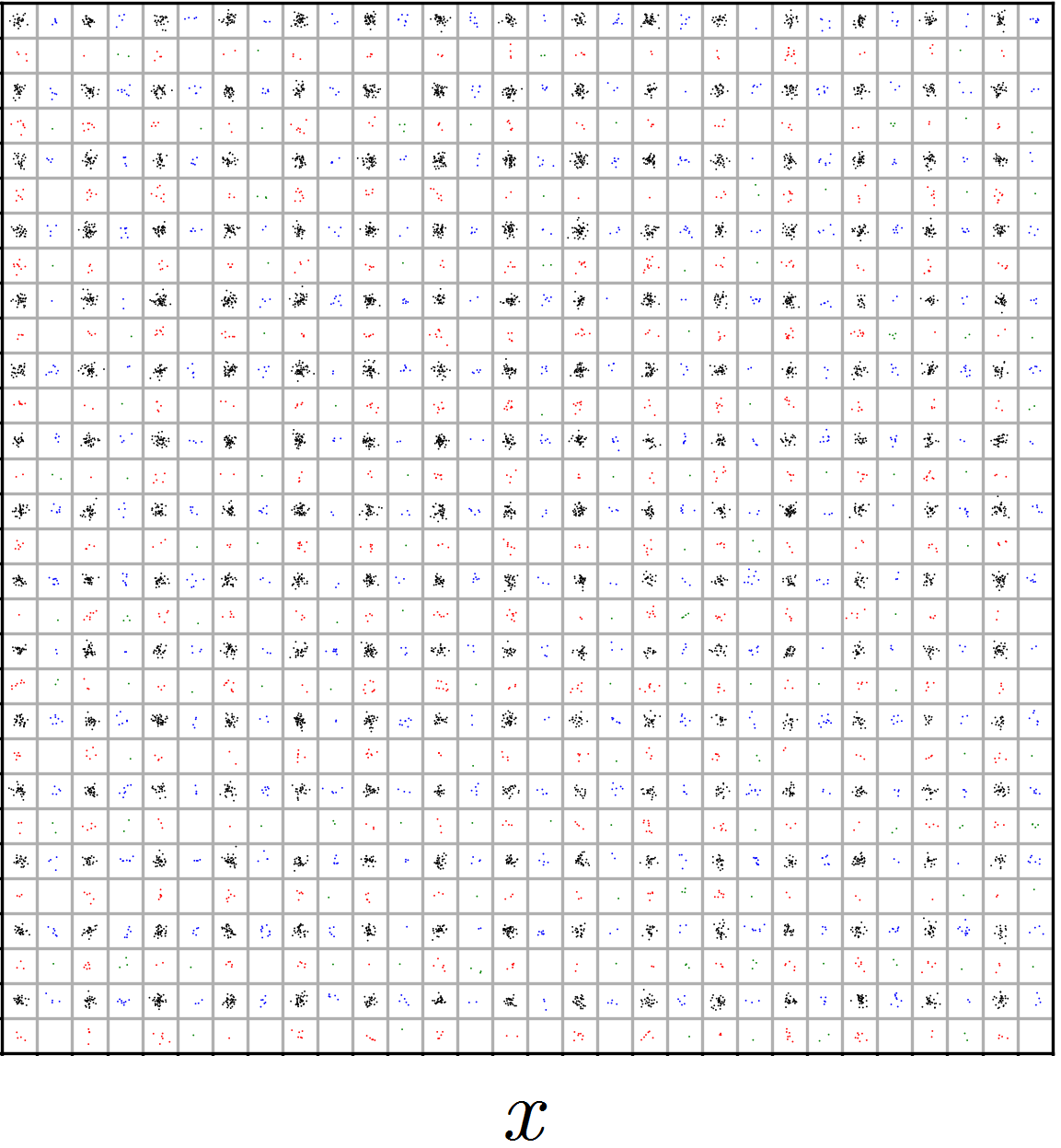}}
	\hfill
	\subfloat[droplet]{\includegraphics[width=0.32\linewidth]{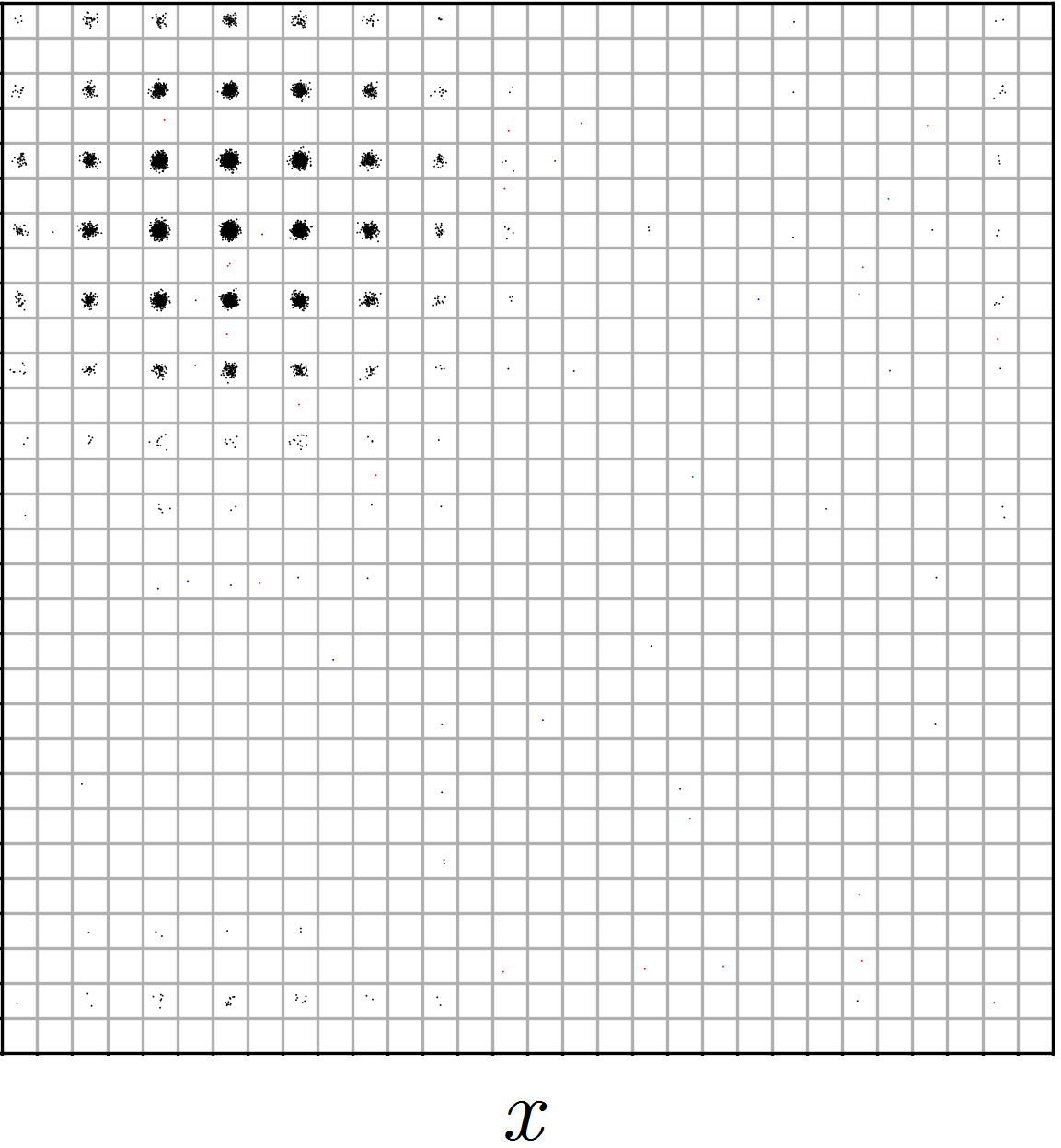}}
	\caption{a) Uniform, b) DW, and c) droplet states for $\mathbb{Z}_2 \times \mathbb{Z}_2$ case for system $30\times30$ with $\xi=7.5$. Sublattice coloring is the same as in Fig. \ref{fig:potential_orthogonal_polarizations}.}
	\label{fig:Z4_modeling_positions}
\end{figure}

\begin{figure}[tbh]
	\centering
	\subfloat[]{\includegraphics[width=0.335\linewidth]{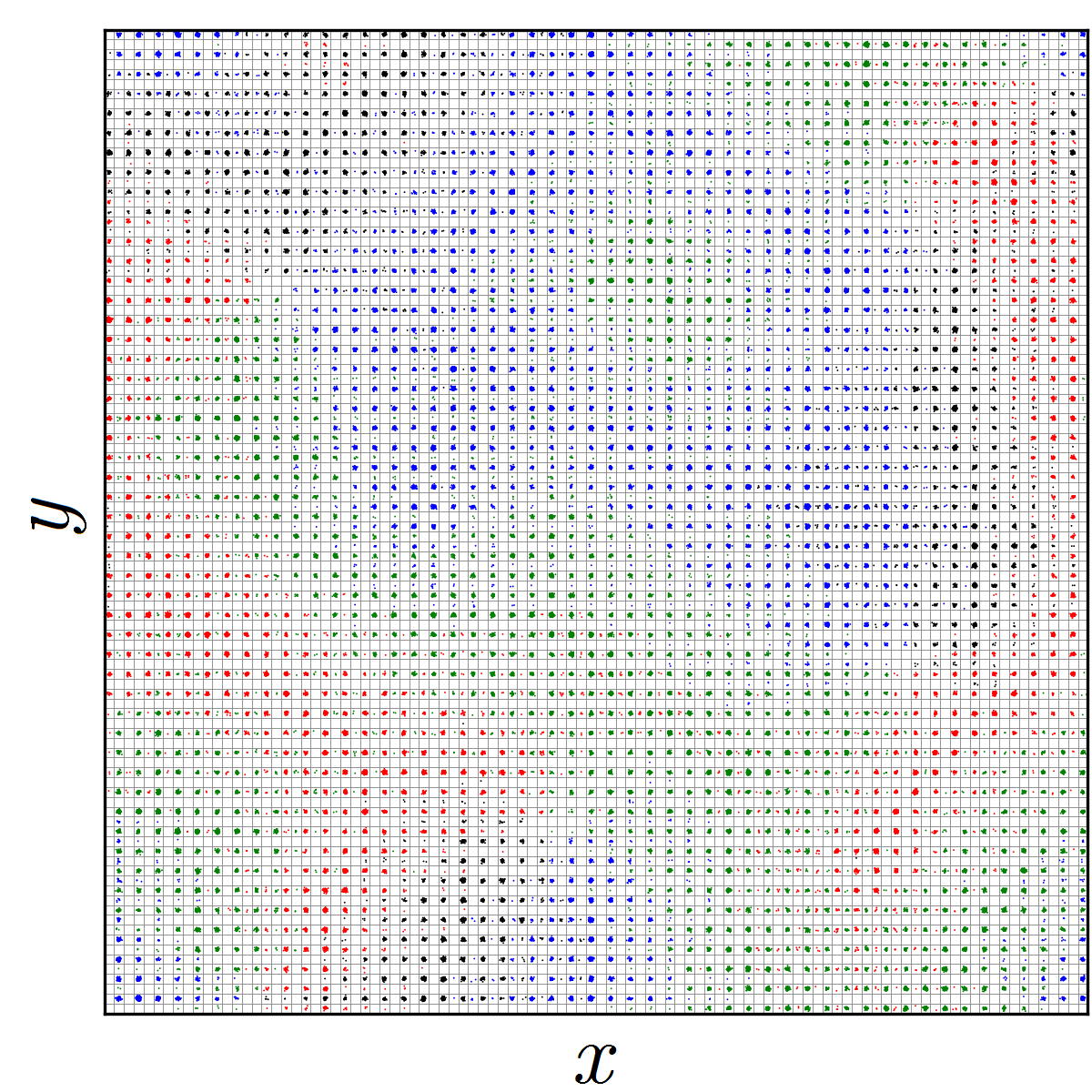}}
	\hfill
	\subfloat[]{\includegraphics[width=0.32\linewidth]{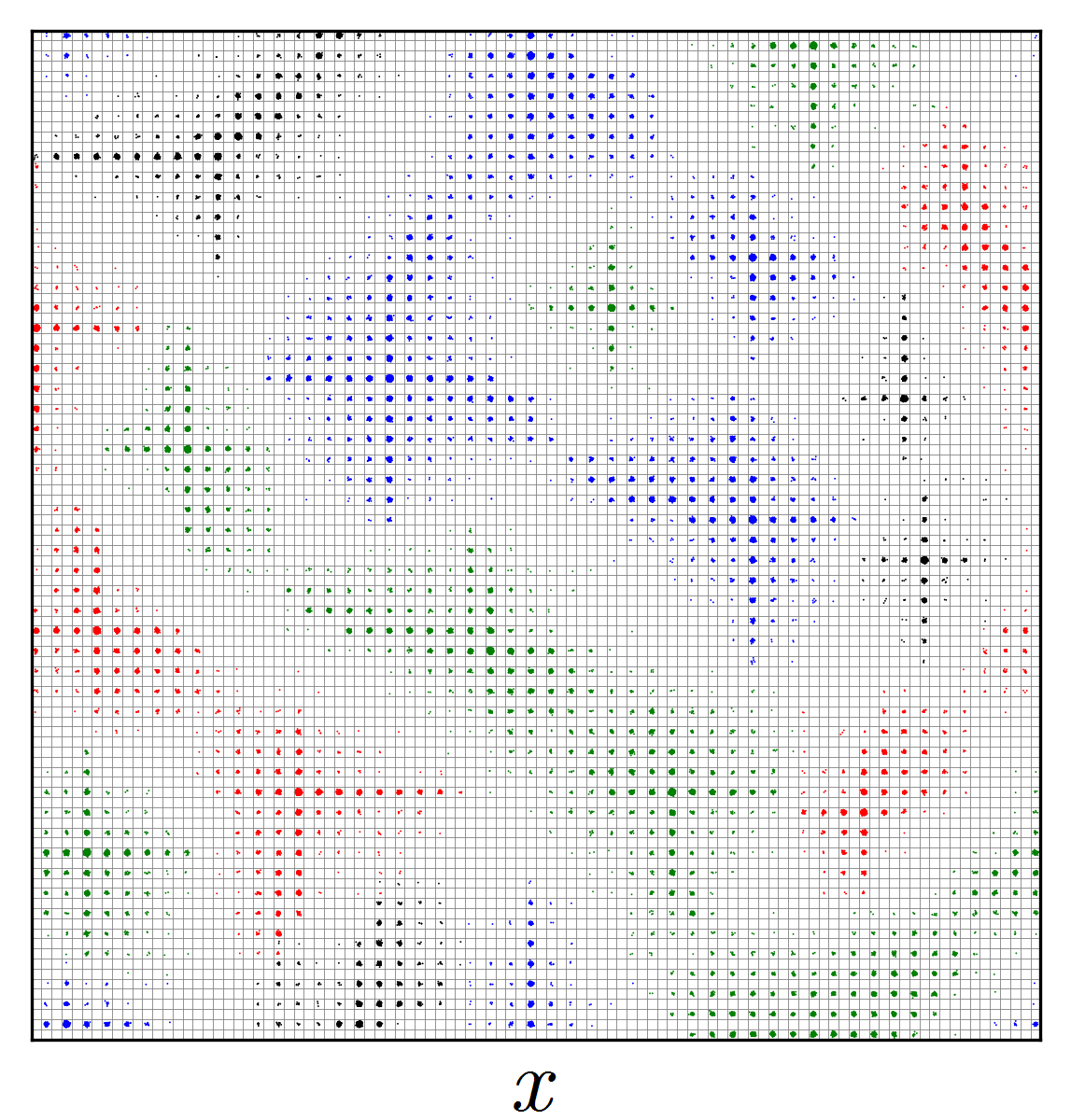}}
	\hfill
	\subfloat[]{\includegraphics[width=0.32\linewidth]{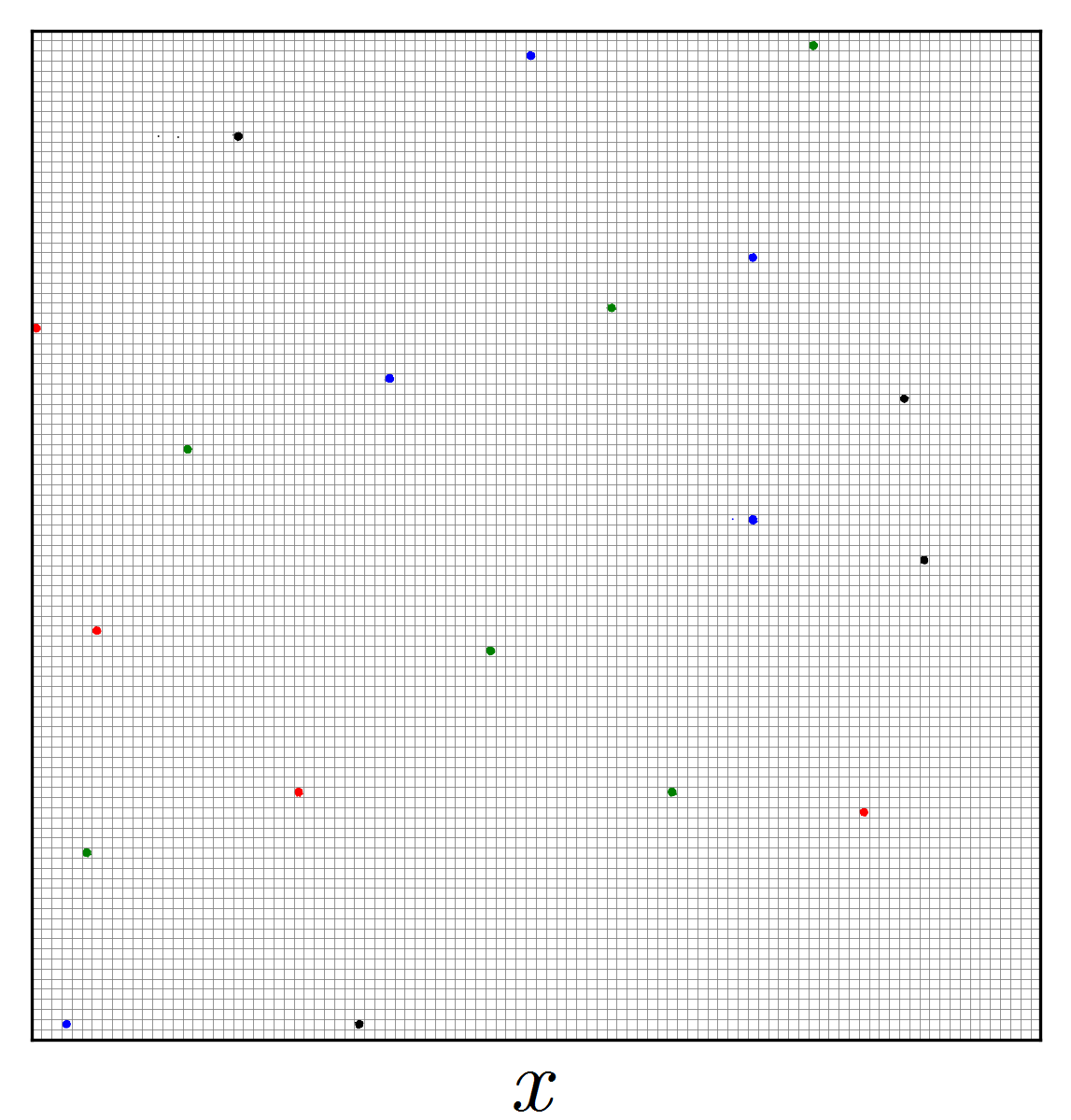}}
	\caption{Monte Carlo evolution of the $100\times100$ system with $\xi=5$, instantaneously quenched from the infinite temperature to $T=U_0$. Red, blue, green, and blacks dots represent particles at four sublattices. a) DW state with several domains (after 10 MC sweeps); b) Domains become sharply separated by regions without particles (after 50 MC sweeps); c) Droplet phase with all the droplets collapsed to one site (after 300 MC sweeps), this state is then preserved for an indefinite time.}
	\label{fig:quench_Z4}
\end{figure}

Figures \ref{fig:Z4_modeling_plots}-\ref{fig:quench_Z4} show the results of the numerical simulations for the $\mathbb{Z}_2 \times \mathbb{Z}_2$ case analogous to Figures 3-5 for $\mathbb{Z}_2$ case presented in the main text.

